\documentclass[12pt,cite,a4paper]{article}

\usepackage{amsmath}
\usepackage{amssymb}
\usepackage[dvips]{graphicx}
\numberwithin{equation}{section}
\renewcommand{\thefootnote}{\fnsymbol{footnote}}

\begin{document}

\begin{titlepage}

\begin{flushright}
\end{flushright}
~

\vskip .8truecm

\begin{center}
\Large\bf
Near-flat space limit and Einstein manifolds
\end{center}

\vskip 1.0truecm

\begin{center}
{Sergio Benvenuti$^1$ \;\;\;and\;\;\; Erik Tonni$^2$} \\
\end{center}

\vskip 1.5truecm

\begin{center}
{\small\it 1. Joseph Henry Laboratories, Princeton University}\\
{\small\it Princeton, NJ 08544, USA}\\
{\small\it e-mail: sbenvenu@princeton.edu}\\
\end{center}
\vskip .8truecm
\begin{center}
{\small\it 2. Scuola Normale Superiore and INFN}\\
{\small\it Piazza dei Cavalieri, 7}\\
{\small\it 56126, Pisa ITALY}\\
{\small\it e-mail: e.tonni@sns.it}\\
\end{center}

\vskip 2.3truecm

\begin{abstract}

\noindent  
We study the near-flat space limit for strings on $AdS_5 \times M^5$, where the internal manifold $M^5$ is equipped with a generic metric with $U(1)^3$ isometry.
In the bosonic sector, the limiting sigma model is similar to the one found for $AdS_5 \times S^5$, as the global symmetries are reduced in the most general case. When $M^5$ is a Sasaki-Einstein space like $T^{1,1}$, $Y^{p,q}$ and $L^{p,q,r}$, whose dual CFT's have $\mathcal{N}=1$ supersymmetry, the near-flat space limit gives the same bosonic sector of the sigma model found for $AdS_5 \times S^5$. 
This indicates the generic presence of integrable subsectors in AdS/CFT.

\end{abstract}

\vskip 1truecm

\noindent

\renewcommand{\thefootnote}{\arabic{footnote}}
\end{titlepage}

\section*{Introduction}

Recent years have seen a deep progress in our understading of the spectrum of maximally supersymmetric AdS/CFT in four dimensions, namely the scaling dimensions of the \mbox{$\mathcal{N} = 4$} SYM and the energies for string states of the type IIB on $AdS_5 \times S^5$ in the planar limit. On both sides of the correspondence, the problem is basically solved in terms of Thermodynamic Bethe Ansatz equations, which allow to compute the spectrum of long states (i.e. the ones with a large $U(1)$ R-charge $J\,$) for any value of the coupling $\lambda$. See for instance \cite{BMN02, GubserKlebanovPolyakov02, MinahanZarembo, BenaRoibanPolchinski02, BeisertStaudacher03,  Kruczenski04, BeisertDippelStaudacher04, ArutyunovFrolovStaudacher04, Beisert04review, Beisert05 SU(2|2), Janik:2006dc, Hernandez:2006tk, HofmanMaldacena06, BeisertHernandezLopez06, BeisertEdenStaudacher06}.

The gauge theory problem is mapped to a spin chain computation. In the spin chain language, one considers a finite set of impurities (magnons) propagating with a definite momentum $p$ along an infinite chain: integrability basically means that multimagnon scattering factorizes into $2 \rightarrow 2$ scatterings and therefore we just need the magnons dispersion relation and the $2$-magnon $S$ matrix to compute the energy of an arbitrary state. Remarkably, we know have an explicit expression for the $S$ matrix \cite{BeisertHernandezLopez06, BeisertEdenStaudacher06}. The classical string theory dual of a magnon (called giant magnon) was found in \cite{HofmanMaldacena06}. In this giant magnon regime the energy $E$ and the spin $J$ are infinite with finite $E-J$, like in the pp-wave limit \cite{BMN02}, but the magnon momentum $p$ is finite and fixed, differently from the pp-wave limit where $p$ is infinitesimal, being $J\sim \sqrt{\lambda}\rightarrow\infty$ with $p\,\sqrt{\lambda}$ fixed.

A particular limit \cite{ArutyunovFrolovStaudacher04, MaldacenaSwanson06} which interpolates between the pp-wave and the giant magnon regimes plays an important role. 
In this ``near-flat space'' limit, $J$ and $\lambda$ go to infinity while $p\,\sqrt[4]{\lambda}$ is kept finite. Maldacena and Swanson  \cite{MaldacenaSwanson06} showed that the worldsheet sigma model drastically simplifies in the near-flat space limit, but the $S$ matrix remains non trivial (differently from the pp-wave limit, where the magnons are free). Interestingly, the $S$ matrix for the near-flat space sigma model of \cite{MaldacenaSwanson06} has been computed up to two loops \cite{KloseZarembo07,KloseMinahanZarembo07} and shown to agree with the near-flat space limit of the full $S$ matrix.

Our aim is to extend some of these important results to four dimensional AdS/CFT dualities with less symmetries. Interesting generalizations of the maximal supersymmetric duality are based on type IIB on backgrounds of the form $AdS_5 \times M^5$, where the internal $M^5$ is a compact Einstein
manifold. In particular, if $M^5$ is Sasaki-Einstein then minimal supersymmetry is preserved. 
In the seminal paper \cite{KlebanovWitten98}, the $\mathcal{N}=1$ gauge theory dual to type IIB on $AdS_5 \times T^{1,1}$ has been found. More recently, infinite families of five dimensional Einstein-Sasaki spaces have been found: $Y^{p,q}$ \cite{GauntlettMartelli} and $L^{p,q,r}$ \cite{CveticLuPagePope05, Martelli:2005wy}. Their gauge theory duals have been constructed respectively in \cite{BenvenutiHananyMartelli04} and \cite{BenvenutiKruczenskiLpqr, Franco:2005sm, ButtiForcellaZaffaroniLpqr}.
Some of the results obtained in $\mathcal{N}=4$ can be extended to these $\mathcal{N}=1$ theories: for instance, in  \cite{BenvenutiKruczenski05Ypq} it has been shown that the classical string theory limit introduced in \cite{Kruczenski04} exists also for a generic Sasaki-Einstein manifold and can be qualitatively connected to a gauge theory spin chain.

After \cite{BMN02}, the Penrose limit has been studied for the other compactifications of the form $AdS_5 \times M^5$ where the explicit metric is known, namely for the spaces $T^{1,1}$ 
\cite{KlebanovItzhaki02,GomisOoguri02,PandoZayasSonnenschein02}, $T^{p,q}$ (which are Einstein but not Sasaki) \cite{KlebanovItzhaki02}, $Y^{p,q}$ and $L^{p,q,r}$ \cite{KupersteinSonnenschein06}. When $M^5$ is a Sasaki-Einstein space, the resulting background is precisely the same one obtained for $S^5$. For $T^{p,q}$ the limiting background has always a pp-wave form, but some of global symmetries are broken.

In this paper we study the near-flat space limit for strings propagating on $AdS_5\times M^5$, taking manifolds like $T^{p,q}$, $Y^{p,q}$ and $L^{p,q,r}$ as internal five dimensional space $M^5$. Since the covariant action for the type IIB superstring on $AdS_5\times M^5$ is known only when $M^5=S^5$ \cite{MetsaevTseytlin98}, we consider only the bosonic sector, namely the Polyakov action. Our result is that for Sasaki-Einstein $M^5$ the near-flat space sigma model is identical to the one found for $S^5$ in \cite{MaldacenaSwanson06}. We show this explicitly for all the known Sasaki-Einstein metrics.
Since the sigma model of \cite{MaldacenaSwanson06} is integrable,
this indicates that the four dimensional $\mathcal{N}=1$ SCFT's with a ten dimensional gravity dual possess an integrable subsector.

The paper is organized as follows.
In section \ref{NFS for Tpq} we study the near-flat space limit for $AdS_5\times T^{p,q}$ backgrounds, obtaining a two dimensional sigma model similar to the bosonic sector of the one found for $AdS_5\times S^5$  \cite{MaldacenaSwanson06}. The special case of $T^{1,1}$, the only one which is
stable (in the sense of Breitenlohner-Freedman \cite{BreitenlohnerFreedman, GubserMitra02}) and supersymmetric, gives a limiting sigma model identical to the $S^5$ case.
In section \ref{NFS Ypq} we consider the near-flat space limit for internal $Y^{p,q}$, recovering again the bosonic sector of the near-flat space sigma model  of type IIB on $AdS_5\times S^5$. Since this situation resembles the one occurring for the Penrose limit, where different geometries give the same limiting result, in section \ref{section U3} we introduce a generalized metric with $U(1)^3$ symmetry and study its Penrose limit (subsection \ref{section penrose U3}) and its near-flat space limit (subsection \ref{section NFS U3}). The limiting sigma model is similar to the one found for $AdS_5 \times S^5$, but the global symmetries are reduced in the most general case. Moreover, we find that  the coefficients characterizing the needed field redefinitions are the same occurring in the coordinate transformations of the Penrose limit.

In appendix \ref{S5} we review the initial steps of the near-flat space limit for  $AdS_5\times S^5$ \cite{MaldacenaSwanson06}, while in appendix \ref{appendix special cases} we apply the considerations made in section  \ref{section U3} to generalized metrics with $U(1)^3$ symmetry which include the known cases in the usual coordinates more directly. We conclude by applying these results also to the special case of $AdS_5\times L^{p,q,r}$.

\section{The near-flat space limit for the $T^{p,q}$ metrics}
\label{NFS for Tpq}

In this section we construct the near-flat space limit of bosonic strings moving in backgrounds of the form $AdS_5 \times T^{p,q}$, studying the Polyakov action and the Virasoro constraints.\\
As is well known, the bosonic sector of closed strings propagating in a ten dimensional target space with metric $G_{MN}$ is described by the Polyakov action
\footnote{We recall that the coupling constant $g_{\scriptscriptstyle \textrm{YM}}$ of the gauge theory is related to the radius $R$ of $AdS_5$ by $\lambda=R^4/\alpha'^{\,2}$, where the 't Hooft coupling $\lambda \equiv g^2_{\scriptscriptstyle \textrm{YM}} N$ is kept fixed in the large $N$ limit. If $M^5$ is not $S^5$, $g_{\scriptscriptstyle \textrm{YM}}$ has to be thought as an overall potential coupling, as discussed in detail in the supersymmetric case in \cite{BH2005,BenvenutiKruczenski05Ypq}.}
\begin{equation}
\label{polyakovaction}
S\,=\,-\,\frac{R^2}{2} \int d\sigma_0\int_0^{2\pi}\frac{d\sigma_1}{2\pi}\,
\,\sqrt{-\gamma}\,\gamma^{ab}\,
G_{MN}\,\partial_a X^M \partial_b X^N
\end{equation}
where $\gamma_{ab}$ is the worldsheet metric, playing the role of a Lagrange multiplier.
The energy momentum tensor
\begin{equation}
\label{emtensor}
T_{ab}\,=\,-\,\frac{4\pi}{R^2}\,\frac{1}{\sqrt{-\gamma}}\,\frac{\delta\,S}{\delta\gamma^{ab}}
\,=\,
G_{MN}\,\partial_a X^M \partial_b X^N
-\frac{1}{2}\,\gamma_{ab}\,\gamma^{cd}\,
G_{MN}\,\partial_c X^M \partial_d X^N
\end{equation}
is symmetric and traceless, and the equations of motion for the worldsheet metric $\gamma_{ab}$ are $T_{ab}=0$. Adopting the conformal gauge $\gamma_{ab}=\eta_{ab}=\textrm{diag}(-1,1)$ and  
introducing the rescaled light-cone worldsheet coordinates $\sigma^\pm$ as follows \cite{MaldacenaSwanson06}
\begin{equation}
\sigma_0+\sigma_1 = 2\sqrt{g}\,\sigma^+\;,
\hspace{1.3cm}
\sigma_0 - \sigma_1= \frac{\sigma^-}{2\sqrt{g}}\;,
\hspace{1cm}
\textrm{where}
\hspace{1cm}
g\,\equiv\,\frac{R^2}{4\pi}\;,
\end{equation}
the Polyakov action becomes
\begin{equation}
\label{stringaction}
S\,=\,-\,2\,g\int   G_{MN} \, \partial_+ X^M \partial_- X^N\;
d\sigma^+ d\sigma^- \;,
\end{equation}
while the components of the energy momentum tensor read
\begin{equation}
\label{Virasoro}
T_{--}
\,=\,G_{MN}\,\partial_- X^M \partial_- X^N\;,
\hspace{1.5cm}
T_{++}
\,=\,G_{MN}\,\partial_+ X^M \partial_+ X^N\;.
\end{equation}
The ten dimensional target spaces we will consider are of the form $AdS_5 \times M^5$, 
where $M^5$ is a compact manifold, so that their metrics are of the form \cite{FreundRubin}
\begin{equation}
\label{tendimmetric}
ds^2\,=\, R^2 \,G_{MN}\, dX^M dX^N\,=\, ds^2_{AdS_5}+ds^2_{M^5}\;.
\end{equation}
The global $AdS_5$ metric reads
\begin{equation}
\label{AdS5metric}
ds^2_{AdS_5}\,=\,R^2\big(-\cosh^2\rho\;dt^2+d\rho^2+\sinh^2\rho\;d\Omega_3^2\,\big)\;,
\end{equation}
where $d\Omega_3^2$ is the metric on $S^3$, which can be written as $d\Omega_3^2\,=\,d\delta_1^2
+\cos^2 \delta_1\,d\delta_2^2+\sin^2 \delta_1\,d\delta_3^2$.

The compact manifolds $T^{p,q}$ are characterized by the metrics \cite{CandelasOssa89}
\begin{equation}
\label{abcmetric}
ds^2_{M^5}=R^2\big[\,a^2(d\psi+p \cos\theta_1 d\phi_1+q \cos\theta_2 d\phi_2)^2
+b^2(d\theta_1^2+\sin^2\theta_1 d\phi_1^2)
+c^2(d\theta_2^2+\sin^2\theta_2 d\phi_2^2)\,\big]\,,
\end{equation}
where the coordinate ranges are $0\leqslant \psi < 4\pi$,  $0\leqslant \theta_i < \pi$, 
$0\leqslant \phi_i < 2\pi$. $p$, $q$, $a^2$, $b^2$ and $c^2$ are parameters. The Einstein condition determines $a^2$, $b^2$ and $c^2$ in terms of the integers $p$ and $q$, but we can keep them unrelated for the moment. In the important special case of $T^{1,1}$ the space is Sasaki-Einstein and the dual CFT is supersymmetric. All the $T^{p,q}$ metrics admit the isometry group $SU(2) \times SU(2)\times U(1)$.\\
We consider the NFS limit around the geodesic sitting at $\theta_1\,=\,\theta_2\,=\,0$. The same geodesics was the starting point for the Penrose limit in \cite{KlebanovItzhaki02}.
Starting from the string action (\ref{stringaction})  we perform the field redefinitions
\begin{eqnarray}
\label{field redefinitions Tpq}
& & t\,=\,k_t \sqrt{g}\,\sigma^{+}+\,\frac{\tau}{\sqrt{g}}\;,
\hspace{4.2cm}\rho\,=\,\frac{z}{\sqrt{g}}\;,\nonumber\\
\rule{0pt}{.76cm}
& & \psi\,=\,k_\psi \sqrt{g}\,\sigma^{+}+\,\frac{K_\psi\,\chi}{K\sqrt{g}}\,- p\,\varphi_1-q\,\varphi_2\;,\\
\rule{0pt}{.76cm}
& & \phi_1\,=\,k_{\phi_1} \sqrt{g}\,\sigma^{+}+\,\frac{K_{\phi_1}\,\chi}{K\sqrt{g}}\,+\,\varphi_1\;,
\hspace{2cm}\theta_1\,=\,\frac{r_1}{b\,\sqrt{g}}\;,
\nonumber\\
\rule{0pt}{.76cm}
& & \phi_2\,=\,k_{\phi_2} \sqrt{g}\,\sigma^{+}+\,\frac{K_{\phi_2}\,\chi}{K\sqrt{g}}\,+\,\varphi_2\;,
\hspace{2cm}\theta_2\,=\,\frac{r_2}{c\,\sqrt{g}}\;,\nonumber
\end{eqnarray}
where $K=a(K_\psi +p\,K_{\phi_1} +q\,K_{\phi_2})$, the $k$'s and $K$'s are constant.\\
Now we substitute the field redefinitions (\ref{field redefinitions Tpq}) into the string action (\ref{stringaction}) with $G_{MN}$ given by (\ref{tendimmetric}) and (\ref{abcmetric}), and take the limit $g\rightarrow \infty$. The term $O(g)$ in the Lagrangian does not contribute to the action because it is a total derivative. Instead, the term  $O(\sqrt{g}\,)$ is proportional to
\begin{eqnarray}
& & \hspace{-.8cm}\sqrt{g}\,\left\{\,
r_1^2\,\partial_{-}\varphi_1
\left(\,\frac{a^2 p}{2\,b^2}\,\big(k_{\psi}+p\,k_{\phi_1}+q\,k_{\phi_2}\big)- k_{\phi_1} \right)
\right.\\
& & \hspace{6.4cm}\left.+\,
r_2^2\,\partial_{-}\varphi_2
\left(\,\frac{a^2 q}{2\,c^2}\,\big(k_{\psi}+p\,k_{\phi_1}+q\,k_{\phi_2}\big)- k_{\phi_2} \right)
\,\right\}\;.\nonumber
\end{eqnarray}
This divergence of the action vanishes provided that 
\begin{equation}
\label{kvaluesTpq}
k_\psi\,=\,\left(1-\,\frac{a^2 p^2}{2\,b^2}\,-\,\frac{a^2 q^2}{2\,c^2}\,\right)\frac{k_\Psi}{a}\;,
\hspace{.7cm}
k_{\phi_1}\,=\,\frac{a\,p}{2\,b^2}\,k_\Psi\;,
\hspace{.7cm}
k_{\phi_2}\,=\,\frac{a\,q}{2\,c^2}\,k_\Psi\;.
\end{equation}
These relations imply for $k_\Psi$
\begin{equation}
\label{kPsi}
k_\Psi\,=\,a\,(k_\psi +p\, k_{\phi_1} +q\, k_{\phi_2})
\end{equation}
for $k_\Psi$, which remains a free parameter.
Adopting (\ref{kvaluesTpq}) into (\ref{field redefinitions Tpq}), we find
\begin{eqnarray}
\label{NFSlagrangianTpq}
\lim_{g\,\rightarrow\,\infty}\,S  &=&
-\,2\int \bigg\{-\,\partial_{+}\tau\,\partial_{-}\tau
+ \partial_{+}\chi\,\partial_{-}\chi
+ \partial_{+}\vec{z}\, \partial_{-}\vec{z}
+\sum_{i\,=1,2}\partial_{+}\vec{r}_i\,\partial_{-}\vec{r}_i
\phantom{\frac{c^2p^2}{b^2}}
\\
\rule{0pt}{.76cm}
& &\hspace{3.7cm}
-\,k_t\,z^2\,\partial_{-}\tau
-\, k_\Psi \left(\,\frac{a^2p^2}{4\,b^4}\,r_1^2
+\,\frac{a^2q^2}{4\,c^4}\,r_2^2\right)\partial_{-}\chi\,\bigg\}\,d\sigma^+ d\sigma^-.\nonumber
\end{eqnarray}
where 
and $\partial_{+}\vec{r}_i\,\partial_{-}\vec{r}_i\,=\,
\partial_{+}r_i\,\partial_{-}r_i+r_i^2\,\partial_{+}\varphi_i\,\partial_{-}\varphi_i$ for $i=1,2$.
As in \cite{MaldacenaSwanson06}, the action is right conformal invariant ($\sigma^- \rightarrow f(\sigma^-)$ with arbitrary $f$), but it is not invariant under left conformal transformations ($\sigma^+ \rightarrow f(\sigma^+)$).\\
Now we turn to the Virasoro constraints. 
Considering first $T_{--}$, one finds that, given the field redefinitions (\ref{field redefinitions Tpq}), the first term of its expansion at large $g$ is 
\begin{equation}
\label{T--final Tpq}
T_{--}=\,\frac{1}{g}\left(-\,(\partial_{-}\tau)^2+\,(\partial_{-}\chi)^2
+ \,(\partial_{-}\vec{z}\,)^2
+\sum_{i\,=\,1}^{2} (\partial_{-}\vec{r}_i)^2\right)
+\,O\big(\,g^{-3/2}\,\big)\;,
\end{equation}
where $(\partial_{-}\vec{z}\,)^2\,=\,
(\partial_{-}z)^2+z^2
\big((\partial_{-}\delta_1)^2
+\cos^2 \delta_1\,(\partial_{-}\delta_2)^2
+\sin^2 \delta_1\,(\partial_{-}\delta_3)^2\big)$
and $(\partial_{-}\vec{r}_i)^2\,=\,
(\partial_{-}r_i)^2+r_i^2\,(\partial_{-}\varphi_i)^2$ for $i=1,2$.
Notice that in obtaining (\ref{T--final Tpq}) we do not need the relations (\ref{kvaluesTpq}), and $k_t$ can also be kept arbitrary.\\
As for the component $T_{++}$, we find that the first term of its expansion at large $g$ is $O(g)$, and imposing its vanishing gives
\begin{equation}
\label{k_t Tpq}
k_t^2\,=\,a^2\,(k_\psi +p\, k_{\phi_1} +q\, k_{\phi_2})^2\;.
\end{equation}
Then, choosing the positive root for $k$ and imposing also (\ref{kvaluesTpq}), one finds
\begin{equation}
T_{++}=\,
-\,2\,k_\Psi\,\partial_{+}(\,\tau\,-\,\chi\,)
-\,k_\Psi^2
\left(z^2+\,\frac{a^2p^2}{4\,b^4}\,r_1^2+\,\frac{a^2q^2}{4\,c^4}\,r_2^2\right)
+\,O(1/g)\;.
\end{equation}
Moreover, making use of (\ref{kvaluesTpq}) in the positive root of (\ref{k_t Tpq}) one gets $k_t= k_\Psi$, and this makes it natural to rescale $\sigma^+$, defining $\hat{\sigma}^+=k_\Psi\,\sigma^+$. We thus find that, given (\ref{kvaluesTpq}) and the positive root of (\ref{k_t Tpq}) for the constants occurring in the field redefinitions, the Polyakov action in the near-flat space limit is
\begin{eqnarray}
\label{NFSlagrangianTpqRescaled}
\lim_{g\,\rightarrow\,\infty}\,S  &=&
-\,2\int \bigg\{-\,\partial_{\hat{+}}\tau\,\partial_{-}\tau
+ \partial_{\hat{+}}\chi\,\partial_{-}\chi
+ \partial_{\hat{+}}\vec{z}\, \partial_{-}\vec{z}
+\sum_{i\,=1,2} \partial_{\hat+}\vec{r}_i\,\partial_{-}\vec{r}_i
\\
\rule{0pt}{.76cm}
& &\hspace{4.7cm}
-\,z^2\,\partial_{-}\tau
-\left(\,\frac{a^2p^2}{4\,b^4}\,r_1^2+\,\frac{a^2q^2}{4\,c^4}\,r_2^2\right)\partial_{-}\chi\,\bigg\}\,
d\hat{\sigma}^+ d\sigma^-,\nonumber
\end{eqnarray}
while the Virasoro constraints $T_{--}=0$ and $T_{++}=0$, to the first non trivial order, give rise to the two equations
\begin{eqnarray}
\label{VC--TpqFinal}
& & 
-\,(\partial_{-}\tau)^2+\,(\partial_{-}\chi)^2
+ \,(\partial_{-}\vec{z}\,)^2
+\sum_{i\,=1,2}\,(\partial_{-}\vec{r}_i)^2
\,=\,0\;,\\
\label{VC++TpqFinal}
\rule{0pt}{.75cm}& & 
2\,\partial_{\hat+}(\,\tau\,-\,\chi\,)
+
z^2+\,\frac{a^2p^2}{4\,b^4}\,r_1^2+\,\frac{a^2q^2}{4\,c^4}\,r_2^2
\,=\,0\;.
\end{eqnarray}
For generic $p$ and $q$, the resulting two dimensional bosonic sigma models admit the symmetries $SO(4) \times SO(2)^2$, where the $SO(4)$ factor acts on the $AdS$ coordinates $\vec{z}$, while the two factors $SO(2)$'s act on $\vec{r}_1$ and $\vec{r}_2$.\\
The symmetry is enhanced in the special case of $p = q$, once the Einstein condition is employed. Indeed, the Einstein condition $R_{ab}=\,(\Lambda/R^2) g_{ab}$ for the metrics (\ref{abcmetric}) provides three equations allowing to express $a^2$, $b^2$ and $c^2$ in terms of $\Lambda$, $p$ and $q$. They can be written as
\begin{equation}
\frac{a^2 p^2}{4\,b^4}\,=\,\frac{1}{2\,b^2}\,-\,\frac{\Lambda}{2}\;,
\hspace{1.5cm}
\frac{a^2q^2}{4\,c^4}\,=\,\frac{1}{2\,c^2}\,-\,\frac{\Lambda}{2}\;,
\hspace{1.5cm}
\frac{1}{b^2}\,+\,\frac{1}{c^2}\,=\,3\,\Lambda\;.
\end{equation}
For $p=q$, the first two relations imply $b^2=c^2$ and the $SO(2)^2$ symmetry is enhanced to $SO(4)$. Then, letting also $\Lambda=4$, the ratios involved in (\ref{NFSlagrangianTpqRescaled}) and (\ref{VC++TpqFinal}) become equal to 1 and the limiting sigma model becomes the same as for $AdS_5\times S^5$ \cite{MaldacenaSwanson06}. As shown in \cite{GubserMitra02}, for $p\neq q$ the spaces $T^{p,q}$ are unstable in the sense of Breitenlohner and Freedman \cite{BreitenlohnerFreedman}. In the context of the AdS/CFT correspondence this instability means that such compactifications do not have a unitary field theory dual. Thus, we conclude that for $AdS_5 \times T^{p,p}$ backgrounds with $T^{p,p}$ satisfying the Einstein condition, the Polyakov action and the Virasoro constraints in the near-flat space limit are the same as the ones obtained on $AdS_5\times S^5$ \cite{MaldacenaSwanson06} (see appendix \ref{S5}).\\
The simplest and most important special case belonging to the $T^{p,p}$ family of Einstein spaces is $T^{1,1}$, which is given by (\ref{abcmetric}) with $p=q=1$, implying that $a^2=1/9$ and $b^2=c^2=1/6$. The space $T^{1,1}$ space has $SU(2)\times SU(2)\times U(1)$ symmetry and the corresponding Calabi Yau cone $Y^6$ is the conifold. It was first considered by Klebanov and Witten \cite{KlebanovWitten98} as an example of AdS/CFT correspondence with $\mathcal{N}=1$ dual gauge theory.

To close this section we observe that the relation (\ref{k_t Tpq}) for $k_t$ comes naturally also from the Penrose limit \cite{KlebanovItzhaki02}. Indeed, 
in terms of the fields defined in (\ref{field redefinitions Tpq}) we have 
\begin{equation}
x^-\,\propto\,g\big(t\,-\,a\,(\psi+p\,\phi_1+q\,\phi_2)\big)\,=\,
\big(k_t -a\,(k_\psi +p\,k_{\phi_1} +q\,k_{\phi_2})\big)\,g^{3/2}\,\sigma^+
+\,\sqrt{g}\,(\tau-\chi)\;,
\end{equation}
and requiring this combination to be $O(\sqrt{g})$ when $g\rightarrow \infty$ gives (\ref{k_t Tpq}) .

\section{The near-flat space limit for the $Y^{p,q}$ metrics}
\label{NFS Ypq}

In this section we consider the near-flat space limit for the bosonic sector of $AdS_5 \times Y^{p,q}$, along the lines followed for $AdS_5 \times T^{p,q}$ in the previous section. We will find the near-flat space limit is exactly the same as for $AdS_5 \times S^{5}$. The same conclusion holds for $AdS_5 \times L^{p,q,r}$ as well, but, instead of studying this case explictly, we will find it as a special case of more general metrics with $U(1)^3$ isometries (section \ref{section U3} and appendix \ref{appendix special cases}).\\
The $Y^{p,q}$ Sasaki Einstein metrics in the canonical form are \cite{GauntlettMartelli}
\begin{eqnarray}
\label{Ypqmetric}
ds^2_{M^5} &=& R^2
\left\{\,\frac{1}{9}\,\big[\,d\psi- (1-c\,y)\cos\theta\, d\phi+y\,d\beta\,\big]^2\right.\\
\rule{0pt}{.7cm}& & \hspace{2cm}
\left.+\,\frac{1-c\,y}{6}\,(d\theta^2+\sin^2\theta\,d\phi^2)+
\,\frac{p(y)}{6}\,(d\beta+c\,\cos\theta\,d\phi)^2+\,\frac{dy^2}{6\,p(y)}\,\right\}\,,\nonumber
\end{eqnarray}
where $c$ is a constant, $p(y)$ is a function which can also depend on $c$ and is positive in the interval $[y_1,y_2]$, delimited by two of its zeros.
The ranges of the coordinates are $0\leqslant \psi \leqslant 2\pi$,  $0\leqslant \phi \leqslant 2\pi$, $0\leqslant \alpha \leqslant 2\pi\,l$, $0\leqslant \theta \leqslant \pi$ and  $y_1\leqslant y \leqslant y_2$, where $\alpha=-(\beta+c\,\psi)/6$ and $l=l(q,p)$, with $q<p$ relative prime integers, but we can keep $l$ arbitrary. The isometry group of (\ref{Ypqmetric}) is $SU(2)\times U(1)\times U(1)$.
The Einstein condition for the metrics (\ref{Ypqmetric}) provides the exact expression for $p(y)$, but we shall not need it to arrive at our conclusions.\\
As a starting point we choose the geodesic around which the expansion is performed to sit at a zero of $p(y)$, as was done for the Penrose limit in \cite{KupersteinSonnenschein06}. To study the near-flat space limit, we redefine some of the ten embedding fields as follows
\begin{eqnarray}
\label{field redefinitions Ypq}
& & t\,=\,k_t \sqrt{g}\,\sigma^{+}+\,\frac{\tau}{\sqrt{g}}\;,
\hspace{5.9cm}
\rho\,=\,\frac{z}{\sqrt{g}}\;,\nonumber\\
\rule{0pt}{.76cm}
& & \psi\,=\,k_\psi \sqrt{g}\,\sigma^{+}+\,\frac{K_\psi\,\chi}{K\sqrt{g}}\,-\varphi_1+\,\frac{2\,y_0}{p'(y_0)}\, \varphi_2\;,\\
\rule{0pt}{.76cm}
& & 
\phi\,=\,k_{\phi} \sqrt{g}\,\sigma^{+}+\,\frac{K_{\phi}\,\chi}{K\sqrt{g}}\,- \,\varphi_1\;,
\hspace{4.2cm}
\theta\,=\,\left(\frac{6}{1-c\,y_0}\right)^{\frac{1}{2}}\frac{r_1}{\sqrt{g}}\;,
\nonumber\\
\rule{0pt}{.87cm}
& & 
\beta\,=\,k_{\beta} \sqrt{g}\,\sigma^{+}+\,\frac{K_{\beta}\,\chi}{K\sqrt{g}}\,
+c\,\varphi_1\,-\,\frac{2}{p'(y_0)}\,\varphi_2\;,
\hspace{1.7cm}\,
y\,=\,y_0+\frac{3}{2}\,p'(y_0)\,\frac{r_2^2}{g}\;,
\nonumber
\end{eqnarray}
where $K=(K_\psi-(1-c\,y_0)K_{\phi}+y_0 K_{\beta} )/3$. The point $y_0$ is a zero of $p(y)$, i.e. it is either $y_1$ or $y_2$, and we assume that $p'(y_0)\neq 0$.\\
Substituting (\ref{field redefinitions Ypq}) into the string action (\ref{stringaction}) with the metric (\ref{Ypqmetric}) for $M^5$ and taking the limit $g\rightarrow \infty$, the term $O(g)$ in the Lagrangian
is a total derivative w.r.t. $\sigma^-$, and therefore does not contribute to the action. Instead, as for the previous case, there is a divergent term $O(\sqrt{g}\,)$ in the action whose vanishing allows to fix $k_\psi$, $k_{\phi}$ and $k_{\beta}$ in terms of the free parameter $k_\Psi$ as follows 
\begin{equation}
\label{kvaluesYpq}
k_\psi\,=\,2\left(1+\,\frac{y_0}{p'(y_0)}\right)k_\Psi\;,
\hspace{1cm}
k_\phi\,=\,-\,k_\Psi\;,
\hspace{1cm}
k_\beta\,=\,\left(c\,-\,\frac{2}{p'(y_0)}\right)k_\Psi\;,
\end{equation}
which imply
\begin{equation}
\label{kPsiYpq}
k_\Psi\,=\,\frac{k_\psi -(1-c\,y_0)k_\phi+y_0 k_\beta}{3}\;.
\end{equation}
Using (\ref{kvaluesYpq}), we find
\begin{eqnarray}
\label{NFSYpqHprime}
\lim_{g\,\rightarrow\,\infty}\,S  
&=&
-\,2\int \bigg\{-\,\partial_{+}\tau\,\partial_{-}\tau
+ \partial_{+}\chi\,\partial_{-}\chi
+ \partial_{+}\vec{z}\, \partial_{-}\vec{z}
+\sum_{i\,=1,2}\partial_{+}\vec{r}_i\,\partial_{-}\vec{r}_i
\\
\rule{0pt}{.5cm}
& &\hspace{6cm}
-k_t\,z^2\,\partial_{-}\tau \,
-k_\Psi \big(\,r_1^2 + r_2^2\,\big)\partial_{-}\chi\, \bigg\}\,
d\sigma^+ d\sigma^-,\nonumber
\end{eqnarray}
where $\partial_{+}\vec{r}_i\,\partial_{-}\vec{r}_i=\partial_{+}r_i\,\partial_{-}r_i+r_i^2\,\partial_{+}\varphi_i\,\partial_{-}\varphi_i$ and $ \partial_{+}\vec{z}\, \partial_{-}\vec{z}$ is given in (\ref{zvec}).\\
As for the Virasoro constraints, one finds that, after having introduced the field redefinitions (\ref{field redefinitions Ypq}), the first term in the expansion of $T_{--}$ at large $g$ is given by (\ref{T--final Tpq}),
without employing the relations (\ref{kvaluesYpq}). Instead, the expansion of  $T_{++}$ begins with a term $O(g)$ whose vanishing gives
\begin{equation}
\label{k_t Ypq}
k_t^2\,=\,\frac{\big(k_\psi -(1-c\,y_0)k_\phi+y_0 k_\beta\big)^2}{9}\;.
\end{equation}
Choosing the positive root and imposing also (\ref{kvaluesYpq}) one gets $k_t= k_\Psi$ and 
\begin{equation}
\label{T++Ypq}
T_{++}=\,
-\,2\,k_\Psi\,
\,\partial_{+}(\,\tau\,-\,\chi\,)
-\,k_\Psi^2
\big(\,z^2+\,r_1^2+r_2^2\,\big)
+\,O(1/g)\;.
\end{equation}
Introducing the rescaled coordinate $\hat{\sigma}^+= k_\Psi\,\sigma^+$ on the worldsheet finally yields
\begin{eqnarray}
\label{NFSYpqHprimeFinal}
\lim_{g\,\rightarrow\,\infty}\,S  &=&
-\,2\int \bigg\{-\,\partial_{\hat{+}}\tau\,\partial_{-}\tau
+ \partial_{\hat{+}}\chi\,\partial_{-}\chi
+ \partial_{\hat{+}}\vec{z}\, \partial_{-}\vec{z}
+\sum_{i\,=1,2}\partial_{+}\vec{r}_i\,\partial_{-}\vec{r}_i
\\
\rule{0pt}{.7cm}
& &\hspace{6.9cm}
-\,z^2\,\partial_{-}\tau \,
-\big(\,r_1^2+r_2^2\,\big)\partial_{-}\chi\, \bigg\}\,
d\hat{\sigma}^+ d\sigma^-,\nonumber
\end{eqnarray}
while, taking the first term of the expansion at large $g$ of $T_{\pm\pm}$, the Virasoro constraints $T_{--}=0$ and $T_{++}=0$ lead respectively to the following equations
\begin{eqnarray}
\label{NFSYpqT--Final}
\rule{0pt}{.7cm}& & 
-\,(\partial_{-}\tau)^2+\,(\partial_{-}\chi)^2
+ \,(\partial_{-}\vec{z}\,)^2
+\sum_{i\,=1,2}\, (\partial_{-}\vec{r}_i)^2
\,=\,0\\
\label{NFSYpqT++Final}
\rule{0pt}{.5cm}& & 
2\,\partial_{\hat{+}}(\,\tau\,-\,\chi\,)
+\,z^2+\,r_1^2+r_2^2\,
\,=\,0\;.
\end{eqnarray}
The expressions for the Polyakov action and for the Virasoro constraints are equal to those obtained for $AdS_5 \times S^5$ by Maldacena and Swanson \cite{MaldacenaSwanson06}.\\
We remark that in finding these results we have assumed nothing about $p(y)$ except that $p(y_0)=0$ and $p'(y_0)\neq0$.
As in the previous case, the relation (\ref{k_t Ypq}) for $k_t$ comes naturally also from the Penrose limit by requiring that the target space coordinate $x^-\propto g\,[\,t-(\psi-(1-c\,y_0)\phi+y_0\beta)\,]/3=O(\sqrt{g}\,)$ when $g\rightarrow \infty$.\\
The explicit function $p(y)$ for $Y^{p,q}$ obtained imposing the Einstein condition on the metric (\ref{Ypqmetric}) is \cite{GauntlettMartelli}
\begin{equation}
\label{Hsolution}
p(y)\,=\,\frac{2\,c\,y^3-3\,y^2+a}{3(1-c\,y)}\;,
\end{equation}
where $a$ is an arbitrary constant. Letting $a=2 \,c\, y_0^3-3 \,y_0^2$, $y_0$ becomes an obvious zero for $p(y)$. One can easily verify that for (\ref{Hsolution}) the following identity holds
\begin{equation}
p'(y)\,=\,-\,2\,y+\frac{c\,p(y)}{1-c\,y}\;,
\end{equation}
which implies that $p'(y_0)=-\,2\,y_0$ when $y_0$ is a zero for $p(y)$.\footnote{It is well known that $T^{1,1}$ is a special cases of $Y^{p,q}$ \cite{GauntlettMartelli}, therefore it is instructive to recover the field redefinitions for $T^{1,1}$ from the ones for $Y^{p,q}$. Taking $c=0$ and $p(y)=1-y^2$ in (\ref{Ypqmetric}) and changing the coordinates as $\theta=\theta_1$, $\phi=-\,\phi_1$, $y=\cos\theta_2$ and $\beta=\phi_2$ one gets the $T^{1,1}$ metric. Notice that $\theta_{2,0}=0$ means $y_0=1$. The expression for $p(y)=1-y^2$ arises from plugging $c=0$ and $a=3$ into (\ref{Hsolution}), therefore here we have already used the Einstein condition. Anyway, to recover the field redefinitions for $T^{1,1}$ we just need the local property $p'(y_0)=-\,2\,y_0$ and not the full expression for $p(y)$: letting $c=0$, $p'(y_0)=-\,2\,y_0$ and $y_0=1$ into (\ref{field redefinitions Ypq}) and taking into account of the change of coordinates between $Y^{p,q}$ and $T^{1,1}$, we get (\ref{field redefinitions Tpq})  specialized for the $T^{1,1}$ parameters. In particular, since $y=\cos\theta_2$, one understands why $y-1$ scales as $1/g$ in (\ref{field redefinitions Ypq})  and $\theta_2$ as $1/\sqrt{g}$ in (\ref{field redefinitions Tpq}).}

\section{Large radius limits for metrics with $U(1)^3$ symmetry}
\label{section U3}

Among all the examples for $M^5$ considered so far, the ones which are also stable Einstein manifolds  give the same results for the Polyakov action and the Virasoro constraints in the near-flat space limit. This remarkable fact happens also for the Penrose limit of the same backgrounds:
one always gets a ten dimensional pp-wave metric of the form \cite{ppwave1, ppwave2, ppwave3}
\begin{equation}
\label{ppwavemetric}
ds^2\,=\,-\,4\,dx^+ dx^-
+\left(\,\sum_{i,j=1}^8 A_{ij}\,x^i x^j\right)(dx^+)^2+\sum_{i=1}^8 dx^i dx^i\;,
\end{equation}
where the matrix $A_{ij}$ is constant. This metric describes the so called Cahen-Wallach spaces \cite{CahenWallach}. In the case of Sasaki-Einstein spaces the matrix $A_{ij}$ is proportional to the identity (preserving an $SO(8)$ symmetry)
\cite{KlebanovItzhaki02,GomisOoguri02,PandoZayasSonnenschein02, KupersteinSonnenschein06}
, while for $T^{p,q}$ with $p \neq q$ the symmetry is broken to $SO(4) \times SO(2)^2$
\cite{KlebanovItzhaki02}.

In the remaining part of the paper, we study these two large radius limits trying to be as general as possible, without relying on a particulare metric. We find that the coefficients characterizing the field redefinitions of the near-flat space limit are the same occurring in the coordinate transformations of the Penrose limit. \\
\noindent In subsection \ref{metricU13} we define the metrics we are going to consider, which always preserve a $U(1)^3$ isometry, and the geodesic that we take as a starting point for the large radius limits. In subsection \ref{section penrose U3} we consider the Penrose limit for these metrics, which includes all the metrics mentioned above as special cases after a change of coordinates, and in subsection \ref{section NFS U3} we study its near-flat space limit.

\subsection{The metrics and the extremal geodesic}
\label{metricU13}

Any five dimensional Sasaki-Einstein metric on the compact manifold $M^5$ can be written as 
\begin{equation}
\label{metricEinstein}
ds^2_{SE} \,=\,\frac{1}{9}\,\big(\,d\psi+A_i\,dx^i\, \big)^2
+\,\frac{1}{6}\,\hat{g}_{ij}\,dx^i dx^j\,\;,
\end{equation}
where $i,j=1,\dots,4$ and with $A_i$ and $\hat{g}_{ij}$ depending on the four coordinates $x^i$, and $\hat{g}_{ij}$ is locally a Kahler-Einstein metric. The shift of the angle $\psi$ are related to the $U(1)_R$ symmetry in the dual SCFT. If there is an additional $U(1)^2$ symmetry (as for all the Einstein metrics which are explicitly known) the metric can be further simplified. We consider a general metric of the following form:
\begin{equation}
\label{metricsigma}
ds^2_{M^5} \,=\,R^2\Big[\,g_{ab} (\vec{\theta}\,)\,d\psi_a \,d\psi_b
\,+\, g_{44} (\vec{\theta}\,)\,d\theta_1^2 
\,+\, g_{55} (\vec{\theta}\,)\,d\theta_2^2\,\Big]\;.
\end{equation}
The non trivial dependence is only on the coordinates $\vec{\theta}^{\textrm{t}}$ and we can arrange the three remaining ones into a vector  $\vec{\psi}^{\textrm{t}}=(\psi_1,\psi_2,\psi_3)=(\psi,\phi_1,\phi_2)$. We find it convenient to introduce also $\tilde{g}_{44} (\vec{\theta}\,) \equiv g_{44} (\vec{\theta}\,)^{-1}$ and $\tilde{g}_{55} (\vec{\theta}\,) \equiv g_{55} (\vec{\theta}\,)^{-1}$. The metric (\ref{metricsigma}) preserves the $U(1)^3$ symmetry given by the shifts of the angles $\psi_a$, and, in general, it does not satisfy neither the Einstein nor the Sasaki condition. We can think of (\ref{metricsigma}) as a $U(1)^3$ fibration over a two dimensional polygon parameterized by the coordinates $\theta_{i}$, see the figure. One of the three circles in the $U(1)^3$ fiber shrinks to zero size on the edges of the polygon, so two of them shrink to zero size on the corners of the polygon.\footnote{This is standard in toric geometry, however here we are not necessarily preserving supersymmetry (in other words the metric cone over our $M^5$ does not have to be Kahler).}

\begin{figure}
\begin{center}
\includegraphics[width=10 cm]{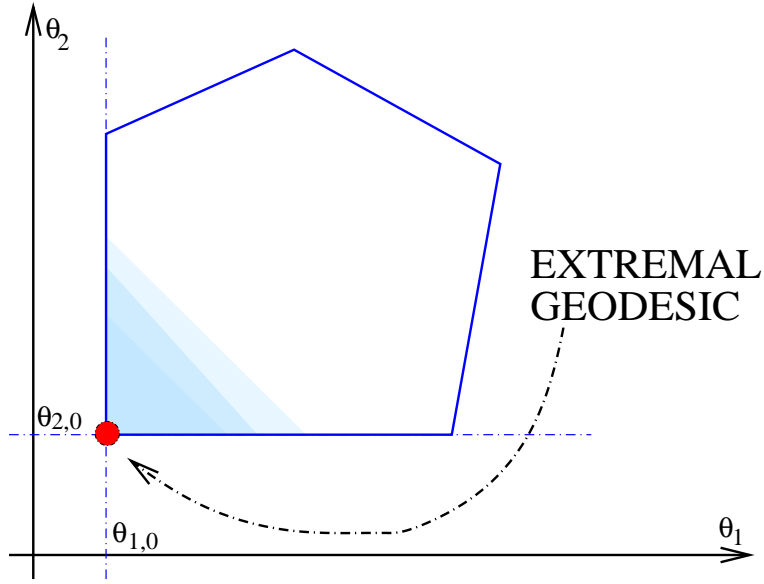}
\caption{Pictorial representation of the two dimensional base of $M^5$. The large radius limits explore only a region of space close to the geodesic.}
\end{center}\label{FIG}
\end{figure}

At this point we have to specify a null geodesic in order to take the large radius limits along it. The null geodesic is sitting at $\rho = 0$ in $AdS_5$ and at $\vec{\theta}^{\textrm{t}}=\vec{\theta}_0^{\textrm{t}}=(\theta_{1,0},\theta_{2,0})$. An important remark is that the values of $\theta_{i,0}$ we will consider are at an \emph{extremal} point: the range of the $\theta_i$ coordinates is $\theta_{i,0} \leq \theta_{i}$, for values of $\theta_{i}$ close to $\theta_{i,0}$. In other words, our geodesic is sitting on a corner of the polygon parameterized by the coordinates $\theta_{i}$, as depicted in the figure. These special geodesics were called \emph{extremal geodesics} in \cite{BenvenutiKruczenski05Ypq}. In the Sasaki-Einstein case, the BPS operators dual to a pointlike string moving along such geodesics are particularly simple to study \cite{BenvenutiKruczenski05Ypq, BenvenutiKruczenskiLpqr}.\footnote{It might be interesting to consider the large radius limits in the case of non-extremal geodesics.}\\
Since $\theta_{i,0}$ is an extremal point we take the functions $\tilde{g}_{44}(\vec{\theta}\,)$ and $\tilde{g}_{55}(\vec{\theta}\,)$ vanishing at $\theta_{i}=\theta_{i,0}$. This means that  $g_{44} (\vec{\theta})$ and $g_{55} (\vec{\theta})$ are both divergent when $\vec{\theta}\rightarrow \vec{\theta}_0$. At $\theta_{i}=\theta_{i,0}$, two out of the three circles in the $U(1)^3$ fibration shrink to zero size and the term $g_{ab} (\vec{\theta}\,)\,d\psi_a \,d\psi_b$ becomes a perfect square $(\sum_I g_{I} \,d\psi_I )^2$; in other words, the following relation holds
\begin{equation}
\label{U3condition3}
g_{ab}(\vec{\theta}_0)\,=\,\big(g_{aa}(\vec{\theta}_0)\,g_{bb}(\vec{\theta}_0)\big)^{\frac{1}{2}}
\hspace{2cm}
a\neq b\;.
\end{equation}
We remark that all the known five dimensional Einstein metrics are special cases of (\ref{metricsigma}).

\subsection{The Penrose limit for a generalized $U(1)^3$ metric}
\label{section penrose U3}

In order to study the Penrose limit of (\ref{metricsigma}), let us consider a null geodesic specified by the values $\rho = 0$ and $\vec{\theta}^{\textrm{t}}=\vec{\theta}_0^{\textrm{t}}=(\theta_{1,0},\theta_{2,0})$. Introducing $\rho=z/R$ and expanding $g_{ab} (\vec{\theta}\,)$ around $\vec{\theta}_0$, the ten dimensional metric of $AdS_5\times M^5$ for large $R$, to the order which is relevant for our purposes, is
\begin{eqnarray}
\label{ds2serieU3}
ds^2 &  &\hspace{-.5cm}=\, R^2\left[\,-\,dt^2+g_{ab} (\vec{\theta}_0)\,d\psi_a \,d\psi_b\,\right]
-z^2 dt^2 +d\vec{z}^{\,2} \\
& &\rule{0pt}{.7cm}
+R^2\,\partial_k g_{ab} (\vec{\theta}_0)\,(\theta_k-\theta_{k,0})\,d\psi_a \,d\psi_b
+R^2\,\frac{\partial_k\partial_p g_{ab} (\vec{\theta}_0)}{2}\,
(\theta_k-\theta_{k,0})(\theta_p-\theta_{p,0})\,d\psi_a \,d\psi_b
\nonumber \\
& &\rule{0pt}{.7cm}
+R^2 \,\frac{d\theta_1^2}{\tilde{g}_{44} (\vec{\theta}_0)+\partial_k \tilde{g}_{44}(\vec{\theta}_0)\,(\theta_k-\theta_{k,0})}
+R^2 \,\frac{d\theta_2^2}{\tilde{g}_{55} (\vec{\theta}_0)+\partial_k \tilde{g}_{55}(\vec{\theta}_0)\,(\theta_k-\theta_{k,0})}\;,
\nonumber
\end{eqnarray}
where $\partial_k\equiv \partial_{\theta_k}$. 
We assume that
\begin{equation}
\label{U3condition1}
\partial_2 \tilde{g}_{44}(\vec{\theta}_0)\,=\,
\partial_1 \tilde{g}_{55}(\vec{\theta}_0)\,=\,0
\hspace{.9cm}
\textrm{and}
\hspace{.9cm}
\partial_1 \tilde{g}_{44}(\vec{\theta}_0)\,\neq\,0\;,
\hspace{.7cm}
\partial_2 \tilde{g}_{55}(\vec{\theta}_0)\,\neq\,0\;,
\end{equation}
which lead to the following definitions of the coordinates $r_k$
\begin{equation}
\label{rhozetaU3}
\theta_k-\theta_{k,0}\,=\,\eta_{k,2}\,\frac{r_k^2}{R^2}
\hspace{1.7cm}
k\,=\,1,2
\end{equation}
where the constant $\eta_{k,2}$ are fixed to give coefficient 1 in front of $dr_k^2$ and read
\begin{equation}
\label{etaU3}
\eta_{1,2}\,=\,\frac{\partial_1 \tilde{g}_{44}(\vec{\theta}_0)}{4}\;,
\hspace{2cm}
\eta_{2,2}\,=\,\frac{\partial_2 \tilde{g}_{55}(\vec{\theta}_0)}{4}\;.
\end{equation}
Notice that, because of the definitions (\ref{rhozetaU3}), the term containing $\partial_k\partial_p g_{ab} (\vec{\theta}_0)$ in (\ref{ds2serieU3}) is infinitesimal when $R\rightarrow \infty$. 
The expansion (\ref{ds2serieU3}) then becomes
\begin{equation}
\label{ds2penroseU30}
ds^2 =R^2\left[\,-\,dt^2+g_{ab} (\vec{\theta}_0)\,d\psi_a \,d\psi_b\,\right]
-z^2 dt^2 +d\vec{z}^{\,2}
+\sum_{k=1,2}\left(\,dr^2_k+ \eta_{k,2}^2\,r_k^2\;
d\vec{\psi}^{\textrm{t}} M^{(k)}d\vec{\psi}\;\right)
\end{equation}
up to infinitesimal term, where we have introduced two $3\times 3$ symmetric matrices $M^{(k)}$, whose elements are 
\begin{equation}
\label{M(k)U3}
M^{(k)}_{ab}\,\equiv\,\partial_k g_{ab} (\vec{\theta}_0)
\hspace{2cm}k\,=\,1,2\;.
\end{equation}
Using eq. (\ref{U3condition3})the $O(R^2)$ term in (\ref{ds2penroseU30}) becomes $ R^2\left[\,-\,dt^2+d\Psi^2\,\right]$, where $\Psi$ is defined as follows
\begin{equation}
\label{Psi}
\Psi\,=\,g_{11}(\vec{\theta}_0)^{\frac{1}{2}}\,\psi
+g_{22}(\vec{\theta}_0)^{\frac{1}{2}}\,\phi_1
+g_{33}(\vec{\theta}_0)^{\frac{1}{2}}\,\phi_2\;.
\end{equation}
At this point one introduces the coordinates  $\vec{\varphi}^{\,\textrm{t}}=(\varphi_1,\varphi_2)$ 
\begin{equation}
\label{def lambda and Omega}
\vec{\phi}\,=\,\vec{\lambda}\,\Psi+\Omega\,\vec{\varphi}\;,
\end{equation}
where the vector $\vec{\lambda}$ and the matrix $\Omega$ have to be fixed. Using (\ref{Psi}) to write $\psi$ in terms of $\Psi$ and $\vec{\phi}$, the vector $\vec{\psi}$ becomes
\begin{equation}
\label{shift}
\vec{\psi}
\,=
\left(\begin{array}{c}
\psi\\
\hline
\vec{\phi}
\end{array}
\right)
=
\left(\begin{array}{c}
\omega_0 \Psi-\vec{\omega}^{\,\textrm{t}}\vec{\phi}\\
\hline
\vec{\phi}
\end{array}
\right)
=\,
\omega\,\left[\,
\left(\begin{array}{c}
1 \\
\hline
\vec{\lambda}
\end{array}\right)\Psi+
\left(\begin{array}{c}
0\\
\hline
\Omega\,\vec{\varphi}
\end{array}
\right)\,\right]\;,
\end{equation}
with the constant matrix $\omega$ given by
\begin{equation}
\label{omega matrix} 
\omega\,=\,
\left(\begin{array}{c|c}
\omega_0  & -\,\vec{\omega}^{\,\textrm{t}}\\
\hline
\vec{0} & \textrm{id}_2
\end{array}
\right)\;,
\end{equation}
whose elements can be read from (\ref{Psi}) and are
\begin{equation}
\omega_0\,=\,g_{11}(\vec{\theta}_0)^{-\frac{1}{2}}\;,
\hspace{1.3cm}
\vec{\omega}^{\,\textrm{t}}\,=\,
\big(\,g_{22}(\vec{\theta}_0)^{\frac{1}{2}}g_{11}(\vec{\theta}_0)^{-\frac{1}{2}}
\,,\;
g_{33}(\vec{\theta}_0)^{\frac{1}{2}}g_{11}(\vec{\theta}_0)^{-\frac{1}{2}}\big)\;.
\end{equation}
Given all these definitions, for $k=1,2$ one obtains
\begin{eqnarray}
\label{dpsiMdpsiU3}
d\vec{\psi}^{\textrm{t}} M^{(k)} d\vec{\psi}
& = &
\left(\begin{array}{c|c}
1 &
\vec{\lambda}^{\textrm{t}}
\end{array}\right)
M^{(k)}_{\omega}
\left(\begin{array}{c}
1 \\
\hline
\vec{\lambda}
\end{array}\right)
d\Psi^2
+\,2
\left(\begin{array}{c|c}
0 &
d\vec{\varphi}^{\,\textrm{t}}\,\Omega^{\,\textrm{t}}
\end{array}\right)
M^{(k)}_{\omega}
\left(\begin{array}{c}
1 \\
\hline
\vec{\lambda}
\end{array}\right)
d\Psi\nonumber
\\
\rule{0pt}{.8cm}& & 
+
\left(\begin{array}{c|c}
0 &
d\vec{\varphi}^{\,\textrm{t}}\,\Omega^{\,\textrm{t}}
\end{array}\right)
M^{(k)}_{\omega}
\left(\begin{array}{c}
0 \\
\hline
\Omega\,d\vec{\varphi}
\end{array}\right)\;,
\end{eqnarray}
where we have introduced the symmetric matrix
\begin{equation}
\label{M omega matrix} 
M^{(k)}_{\omega}
\,=\,
\omega^{\textrm{t}}\,M^{(k)}\,\omega\;.
\end{equation}
It is convenient to write $M^{(k)}_{\omega}$ in the form
\begin{equation}
M^{(k)}_\omega
\,=
\left(\begin{array}{c|c}
m^{(k)}_\omega & \vec{\mu}^{(k)\,\textrm{t}}_\omega\\
\hline
\vec{\mu}^{(k)}_\omega & \phantom{\rule{0pt}{.6cm}} H^{(k)}_\omega
\end{array}\right)\;,
\end{equation}
and to express its elements in terms of $\omega_0\,$, $\vec{\omega}$ and $M^{(k)}$ as
\begin{eqnarray}
m^{(k)}_\omega
&=&
\omega_0^2\,m^{(k)}\;,
\\
\vec{\mu}^{(k)}_\omega
&=&
\omega_0 \big(\,\vec{\mu}^{(k)}-m^{(k)}\,\vec{\omega}\,\big)\;,
\\
H^{(k)}_\omega
&=&
H^{(k)}-\vec{\mu}^{(k)}\,\vec{\omega}^{\,\textrm{t}} 
-\vec{\omega}\;\vec{\mu}^{(k)\textrm{t}} 
+m^{(k)}\,\vec{\omega}\;\vec{\omega}^{\,\textrm{t}}\;,
\end{eqnarray}
where the quantities without the index $\omega$ refer to $M^{(k)}$.
In order to obtain a pp-wave metric the term mixing $d\Psi$ and $d\vec{\varphi}^{\,\textrm{t}}$ in (\ref{dpsiMdpsiU3})  must vanish and this condition allows to fix $\vec{\lambda}$ in our change of variables \ref{def lambda and Omega}. Indeed
\begin{equation}
M^{(k)}_{\omega}
\left(\begin{array}{c}
1 \\
\hline
\vec{\lambda}
\end{array}\right)
\,=\,
\left(\begin{array}{c}
\vec{\mu}^{(k)\,\textrm{t}}_\omega \vec{\lambda} + m^{(k)}_\omega \\
\hline
H^{(k)}_\omega \vec{\lambda} + \vec{\mu}^{(k)}_\omega
\end{array}\right)
\end{equation}
and the term mixing $d\Psi$ and $d\vec{\varphi}^{\,\textrm{t}}$ thus vanishes provided that $\vec{\lambda}$ satisfies
\begin{equation}
\label{lambda vec U3}
H^{(k)}_\omega \vec{\lambda} + \vec{\mu}^{(k)}_\omega\,=\,\vec{0}
\hspace{2cm} k\,=\,1,2\;.
\end{equation}
In general $M^{(1)}\neq M^{(2)}$ (which implies $M^{(1)}_\omega\neq M^{(2)}_\omega$) and we want to keep them unrelated, so that both matrices $H^{(k)}_\omega$ must have a vanishing determinant.
Indeed, if one of them were invertible, one would find $\vec{\lambda}$ from the equation associated to it and the remaining one would become a non trivial relation among the elements of $M^{(1)}_\omega$ and $M^{(2)}_\omega$.\\
Thus, given the vector $\vec{\lambda}$ satisfying (\ref{lambda vec U3}), one gets
\begin{equation}
\label{dpsiMdpsiRed}
d\vec{\psi}^{\textrm{t}} M^{(k)} d\vec{\psi}
\,=\,
\left(\,\vec{\mu}^{(k)\,\textrm{t}}_\omega \vec{\lambda} + m^{(k)}_\omega\right)
d\Psi^2
+\,d\vec{\varphi}^{\,\textrm{t}}\,\Omega^{\textrm{t}}\,H^{(k)}_\omega\,\Omega\,d\vec{\varphi}\;.
\end{equation}
Now one introduces the target space coordinates $x^\pm$ in the usual way
\begin{equation}
\label{x^pm}
t\,=\,\mu\,x^+ + \frac{x^-}{\mu\,R^2}\;,
\hspace{2cm}
\Psi\,=\,\mu\,x^+-\frac{x^-}{\mu\,R^2}\;,
\end{equation}
and (\ref{ds2penroseU30}) becomes 
\begin{eqnarray}
\label{ds2penroseU31}
ds^2 & &\hspace{-.5cm}
=\, -\,4\,dx^+ dx^-
+\,\mu^2 \left\{-\,z^2
+\sum_{k=1,2}r_k^2\,\eta_{k,2}^2
\left(\,\vec{\mu}^{(k)\,\textrm{t}}_\omega \vec{\lambda} + m^{(k)}_\omega\right)
\,\right\}\,(dx^+)^2\nonumber\\
\rule{0pt}{.7cm}& &
+\,d\vec{z}^{\,2}
+\sum_{k=1,2}\Big( dr^2_k+\,r_k^2\,\eta_{k,2}^2\,
d\vec{\varphi}^{\,\textrm{t}}\,\Omega^{\textrm{t}}\,H^{(k)}_\omega\,\Omega\,d\vec{\varphi}
\,\Big)\;,
\end{eqnarray}
where $\eta_{k,2}$ are specified in (\ref{etaU3}) and terms $O(1/R)$ have been neglected.\\
Notice that the term multiplying $(dx^+)^2$ in (\ref{ds2penroseU31}) is fixed, being the vector $\vec{\lambda}$ determined as the solution of (\ref{lambda vec U3}), but we are still free to choose the matrix $\Omega$. To fix $\Omega$, we employ the following fact:
if $\textrm{tr}\,H^{(k)}_\omega\neq 0$ and there is no real $\rho\neq 0$ such that $H^{(1)}_\omega=\rho^2\,H^{(2)}_\omega$, one can always find the matrix $\Omega$ satisfying
\begin{equation}
\label{OmegaU3}
\eta_{1,2}^2\,
\Omega^{\textrm{t}}
H^{(1)}_\omega
\Omega\,=\,
\left(\begin{array}{cc}
1 & 0\\
0 & 0
\end{array}\right)
\hspace{.8cm}\textrm{and}\hspace{.8cm}
\eta_{2,2}^2\,
\Omega^{\textrm{t}}
H^{(2)}_\omega
\Omega\,=\,
\left(\begin{array}{cc}
0 & 0\\
0 & 1
\end{array}\right)\;.
\end{equation}
Indeed\footnote{\,We are grateful to Francesco Bonsante for providing us this argument.}, since $\textrm{det}\,H^{(k)}_\omega=0$, the eigenvalues of $H^{(k)}_\omega$ are $0$ and $\textrm{tr}\,H^{(k)}_\omega\neq 0$. This means that the rank of $H^{(k)}_\omega$ is 1 and therefore there is a non trivial vector $w^{(k)}$ such that $H^{(k)}_\omega=w^{(k)}\,w^{(k)\,\textrm{t}}$. The hypothesis $H^{(1)}_\omega\neq\rho^2\,H^{(2)}_\omega$ implies that $w^{(1)}$  and $w^{(2)}$ are linearly independent.
Now let us choose two non trivial vectors $v^{(k)}\in \textrm{ker}\,H^{(k)}_\omega$. Since the kernel of $H^{(k)}_\omega$ is the linear space orthogonal to $w^{(k)}$, from the linear indepedence of $w^{(1)}$  and $w^{(2)}$, one can see that $v^{(1)}$  and $v^{(2)}$ are also linearly independent. Thus, considering $\{v^{(2)},v^{(1)}\}$ as a basis for $\mathbb{R}^2$ and writing the bilinear products given by $H^{(k)}_\omega$ in this basis, one gets
\begin{eqnarray}
\rule{0pt}{.84cm}
\left(\begin{array}{cc}
v^{(2)\textrm{t}} H^{(1)}_\omega v^{(2)} & v^{(2)\textrm{t}} H^{(1)}_\omega v^{(1)}\\
v^{(1)\textrm{t}} H^{(1)}_\omega v^{(2)} & v^{(1)\textrm{t}} H^{(1)}_\omega v^{(1)}
\end{array}\right)
& = &
\left(\begin{array}{cc}
v^{(2)\textrm{t}} H^{(1)}_\omega v^{(2)} & 0\\
0 & 0
\end{array}\right)\;,\nonumber\\
\rule{0pt}{1cm}
\left(\begin{array}{cc}
v^{(2)\textrm{t}} H^{(2)}_\omega v^{(2)} & v^{(2)\textrm{t}} H^{(2)}_\omega v^{(1)}\\
v^{(1)\textrm{t}} H^{(2)}_\omega v^{(2)} & v^{(1)\textrm{t}} H^{(2)}_\omega v^{(1)}
\end{array}\right)
& = &
\left(\begin{array}{cc}
0 & 0\\
0 & v^{(1)\textrm{t}} H^{(2)}_\omega v^{(1)}
\end{array}\right)\;.
\end{eqnarray}
The matrix $\Omega$ changes the basis from the one we are using to $\{v^{(2)},v^{(1)}\}$, and one can always choose $v^{(2)}$  and $v^{(1)}$ so that $v^{(2)\textrm{t}} H^{(1)}_\omega v^{(2)}=\eta_{1,2}^{-2}$ and $v^{(1)\textrm{t}} H^{(2)}_\omega v^{(1)}=\eta_{2,2}^{-2}$.\\
With this choice of $\Omega$, the metric (\ref{ds2penroseU31}) becomes the pp-wave metric 
\begin{equation}
\label{ds2penroseU3final}
ds^2 
=
-\,4\,dx^+ dx^-
+\,\mu^2 \left\{-\,z^2
+\sum_{k=1,2}r_k^2\,\eta_{k,2}^2
\left(\,\vec{\mu}^{(k)\,\textrm{t}}_\omega \vec{\lambda} + m^{(k)}_\omega\right)
\right\}\,(dx^+)^2
+d\vec{z}^{\,2}
+\sum_{k=1,2} d\vec{r}^{\;2}_k\,,
\end{equation}
where $d\vec{r}^{\;2}_k=dr^{2}_k+r_k^2\,d\varphi_k^2$ and the matrix $A_{ij}$ is diagonal (see (\ref{ppwavemetric})).
As already remarked at the beginning of this section, for all the metrics we are considering $g_{11} (\vec{\theta}\,)$ is constant and this implies $m^{(k)}=0$ for $k=1,2$, which slightly simplify (\ref{ds2penroseU3final}).

\subsection{The near-flat space limit for the $U(1)^3$ metric} 
\label{section NFS U3}

In this subsection we consider the near-flat space limit of the Polyakov action and the Virasoro constraints for $AdS_5 \times M^5$, where $M^5$ is equipped with the generalized $U(1)^3$ metric introduced in the previous subsection.\\
We begin by considering the Lagrangian occurring in the Polyakov action (\ref{stringaction})
\begin{equation}
\label{LagrangianNFSU3}
\mathcal{L}\,=\,-\,2\,g \;  G_{MN} \, \partial_+ X^M \partial_- X^N\;,
\end{equation}
where the target space metric $G_{MN}$ describes $AdS_5 \times M^5$ and the metric on the compact $M^5$ can be read from (\ref{metricsigma}). 
Introducing the field $z$ as a rescaling of $\rho=z/\sqrt{g}$, the fields $r_k$  as
\begin{equation}
\label{rhozetaNFSU3}
\theta_k-\theta_{k,0}\,=\,\eta_{k,2}\,\frac{r_k^2}{g}
\hspace{1.7cm}
k\,=\,1,2
\end{equation}
with $\eta_{k,2}$ given by (\ref{etaU3}) and the field $\Psi$ by (\ref{Psi}), the expansion of the Lagrangian (\ref{LagrangianNFSU3}) reads
\begin{eqnarray}
\label{LagrangianNFSU31}
\mathcal{L} & = &
-\,2\,\left\{\,g\,\big[-\partial_{+} t \,\partial_{-} t +\,\partial_{+} \Psi \,\partial_{-} \Psi\,\big]
\phantom{\frac{\eta_{k,1}^2}{2}}\right.
\\
\rule{0pt}{.4cm}& & \hspace{1.1cm}
\left.-\,z^2 \,\partial_{+} t \,\partial_{-} t  + \partial_{+} \vec{z} \,\partial_{-} \vec{z} 
+\sum_{k=1,2}\left(\partial_{+} r_k \,\partial_{-} r_k +r_k^2\,\eta_{k,2}^2\,
\partial_{-}\vec{\psi}^{\,\textrm{t}} M^{(k)}\,\partial_{+}\vec{\psi}\,\right)\right\}\;,
\nonumber
\end{eqnarray}
where the matrices $M^{(k)}$ have been defined in (\ref{M(k)U3}) and terms infinitesimal when $g\rightarrow \infty$ have been neglected.\\
In order to analyze the near-flat space limit, we perform the following field redefinitions
\begin{equation}
\label{t and Psi NFS}
t\,=\,k_t \sqrt{g}\,\sigma^{+}+\,\frac{\tau}{\sqrt{g}}\;,
\hspace{2cm}
\Psi\,=\,k_\Psi \sqrt{g}\,\sigma^{+}+\,\frac{\chi}{\sqrt{g}}\;,
\end{equation}
where $k_t$ and $k_\Psi$ are constants, so that the expansion (\ref{LagrangianTpqsigma1}) becomes
\begin{eqnarray}
\label{LagrangianNFSU32}
\mathcal{L} & = &
-\,2\,\left\{\,g\,\big[-k_t\,\partial_{-} \tau +\,k_\Psi \,\partial_{-} \chi\,\big]
-\partial_{+} \tau \,\partial_{-} \tau +\,\partial_{+} \chi \,\partial_{-} \chi
\phantom{\frac{\eta_{k,1}^2}{2}}\right.
\\
\rule{0pt}{.5cm}& & \hspace{1.1cm}
\left.-\,k_t\,z^2 \,\partial_{-} \tau  + \partial_{+} \vec{z} \,\partial_{-} \vec{z} 
+\sum_{k=1,2}\left(\partial_{+} r_k \,\partial_{-} r_k +r_k^2\,\eta_{k,2}^2\,
\partial_{-}\vec{\psi}^{\,\textrm{t}} M^{(k)}\,\partial_{+}\vec{\psi}\,\right)\right\}\;.
\nonumber
\end{eqnarray}
Notice that the divergent term $O(g)$ is a total derivative, and therefore can be ignored in the limit of the Polyakov action.
Now we redefine $\vec{\phi}^{\,\textrm t}=(\phi_1,\phi_2)$ as
\begin{equation}
\label{phi sigma NFS}
\vec{\phi}\,=\,\vec{k}_\phi\, \sqrt{g}\,\sigma^{+}+\,\frac{\vec{K}_\phi\,\chi}{K\sqrt{g}}\,+\,\Pi\,\vec{\varphi}\;,
\end{equation}
where $K$, the vectors $\vec{k}_\phi^{\;\textrm{t}}=(k_{\phi_1},k_{\phi_2})$ and 
$\vec{K}_\phi^{\,\textrm{t}}=(K_{\phi_1},K_{\phi_2})$ and the matrix $\Pi$ are constant quantities.
Plugging (\ref{phi sigma NFS}) and the second equation of (\ref{t and Psi NFS}) into the definition (\ref{Psi}), one gets the field redefinition for $\psi$, and, as a consequence, also the following relations
\begin{eqnarray}
\label{kPsi U3}
k_\Psi &=& g_{11}(\vec{\theta}_0)^{\frac{1}{2}}\,k_\psi
+g_{22}(\vec{\theta}_0)^{\frac{1}{2}}\,k_{\phi_1}
+g_{33}(\vec{\theta}_0)^{\frac{1}{2}}\,k_{\phi_2}\;,
\\
\rule{0pt}{.6cm}
K &=& g_{11}(\vec{\theta}_0)^{\frac{1}{2}}\,K_\psi
+g_{22}(\vec{\theta}_0)^{\frac{1}{2}}\,K_{\phi_1}
+g_{33}(\vec{\theta}_0)^{\frac{1}{2}}\,K_{\phi_2}\;.
\end{eqnarray}
The vector $\vec{\psi}^{\,\textrm{t}}=(\psi,\phi_1,\phi_2)$ thus reads
\begin{equation}
\vec{\psi}\,=\,
\omega\,\left[\,
\left(\begin{array}{c}
k_\Psi \\
\hline
\vec{k}_\phi
\end{array}\right)\sqrt{g}\,\sigma^+
+
\left(\begin{array}{c}
1 \\
\hline
\vec{K}_\phi/ K
\end{array}\right)\frac{\chi}{\sqrt{g}}
+
\left(\begin{array}{c}
0\\
\hline
\Pi\,\vec{\varphi}
\end{array}
\right)\,\right]\;,
\end{equation}
with the matrix $\omega$ defined in (\ref{omega matrix}).
The terms containing $r_k^2$ in (\ref{LagrangianNFSU32}), when expanded to the relevant order, become
\begin{eqnarray}
\label{dpsiMdpsiNFS}
\partial_{-}\vec{\psi}^{\,\textrm{t}} M^{(k)}\,\partial_{+}\vec{\psi}
& &\hspace{-.5cm} \,=\,
\left(\begin{array}{c|c}
0 &
\partial_{-}\vec{\varphi}^{\,\textrm{t}} \,\Pi^{\,\textrm{t}}
\end{array}\right)
M^{(k)}_{\omega}
\left(\begin{array}{c}
k_\Psi \\
\hline
\vec{k}_\phi
\end{array}\right)
\sqrt{g}\,
+
\left(\begin{array}{c|c}
1 &
\vec{K}_\phi^{\,\textrm{t}}/K
\end{array}\right)
M^{(k)}_{\omega}
\left(\begin{array}{c}
k_\Psi \\
\hline
\vec{k}_\phi
\end{array}\right)\partial_{-}\chi
\nonumber
\\
\rule{0pt}{.9cm}& & 
+
\left(\begin{array}{c|c}
0 &
\partial_{-}\vec{\varphi}^{\,\textrm{t}} \,\Pi^{\,\textrm{t}}
\end{array}\right)
M^{(k)}_{\omega}
\left(\begin{array}{c}
0 \\
\hline
\Pi\,\partial_{+}\vec{\varphi}
\end{array}\right)+\,O(1/\sqrt{g}\,)\;,
\end{eqnarray}
where the matrices $M^{(k)}_{\omega}$ ($k=1,2$) are given in (\ref{M omega matrix}). Now, since
\begin{equation}
M^{(k)}_{\omega}
\left(\begin{array}{c}
k_\Psi \\
\hline
\vec{k}_\phi
\end{array}\right)
\,=\,
\left(\begin{array}{c}
\vec{\mu}^{(k)\,\textrm{t}}_\omega \vec{k}_\phi + m^{(k)}_\omega k_\Psi
\\
\hline
H^{(k)}_\omega  \vec{k}_\phi + \vec{\mu}^{(k)}_\omega k_\Psi
\end{array}\right)\;,
\end{equation}
imposing the vanishing of the divergent term in (\ref{dpsiMdpsiNFS}) provides the two equations
\begin{equation}
\label{k vec U3}
H^{(k)}_\omega  \vec{k}_\phi + \vec{\mu}^{(k)}_\omega k_\Psi \,=\,\vec{0}
\hspace{2cm} k\,=\,1,2\;.
\end{equation}
Comparing them with (\ref{lambda vec U3}), we find
\begin{equation}
\label{kvalues=lambda}
\frac{\vec{k}_\phi}{k_\Psi }\,=\,\vec{\lambda}\;,
\end{equation}
which tells that the change of coordinates occurring in the Penrose limit fixes some parameters of the field redefinitions of the near-flat space limit. Once we know $\vec{k}_\phi$, then $k_\psi$ follows from (\ref{kPsi U3}).
Given $\vec{k}_\phi/k_\Psi$ solving (\ref{k vec U3}), the expansion (\ref{dpsiMdpsiNFS}) becomes
\begin{equation}
\label{dpsiMdpsiRedNFS}
\partial_{-}\vec{\psi}^{\,\textrm{t}} M^{(k)}\,\partial_{+}\vec{\psi}
\,=\,
\left(\, \vec{\mu}^{(k)\,\textrm{t}}_\omega \vec{k}_\phi + m^{(k)}_\omega k_\Psi \right)
\partial_{-}\chi
+\,
\partial_{-}\vec{\varphi}^{\,\textrm{t}} \,\Pi^{\,\textrm{t}}\,H^{(k)}_\omega\,\Pi\,\partial_{+}\vec{\varphi}
\,+O(1/\sqrt{g}\,)\;.
\end{equation}
At this point, choosing
\begin{equation}
\label{Pi=Omega}
\Pi\,=\,\Omega\;,
\end{equation}
precisely the matrix found for the Penrose limit as the solution of (\ref{OmegaU3}), for the Lagrangian in the near-flat space limit one obtains
\begin{eqnarray}
\label{LagrangianTpqsigmaNFS3}
\mathcal{L} & = &
-\,2\,\left\{\,g\,\big[-k_t\,\partial_{-} \tau +\,k_\Psi \,\partial_{-} \chi\,\big]
-\partial_{+} \tau \,\partial_{-} \tau +\,\partial_{+} \chi \,\partial_{-} \chi
+ \partial_{+} \vec{z} \,\partial_{-} \vec{z} 
+\sum_{k=1,2} \partial_{+} \vec{r}_k \,\partial_{-} \vec{r}_k
\rule{0pt}{.8cm}\right.
\nonumber\\
\rule{0pt}{.8cm}& & \hspace{1.1cm}
\left.-\,k_t\,z^2 \,\partial_{-} \tau \, 
+\,k_\Psi \left[\;\sum_{k=1,2} r_k^2\,\eta_{k,2}^2 
\left( \vec{\mu}^{(k)\,\textrm{t}}_\omega \frac{\vec{k}_\phi}{k_\Psi} + m^{(k)}_\omega \right)\,\right]
\partial_{-}\chi\,\right\}\;,
\end{eqnarray}
up to $O(1/\sqrt{g}\,)$ terms, where the coefficients multiplying $r_k^2$ ($k=1,2$) within the square brackets are fixed. The constants $k_t$ and $k_\Psi$ can be related studying one of the Virasoro constraints.\\
Considering $T_{\pm \pm}$ given in (\ref{Virasoro}), an analysis similar to the one performed for the Lagrangian leads to 
\begin{eqnarray}
\label{VirasoroU3}
T_{\pm\pm} & = & 
-\,(\partial_{\pm} t)^2 \,+\,(\partial_{\pm} \Psi)^2 
\phantom{\frac{x}{x}}
\\
\rule{0pt}{.5cm}& & \hspace{0cm}
+\,\frac{1}{g}
\left[\,-\,z^2 \,(\partial_{\pm} t)^2  + (\partial_{\pm}\vec{z}\,)^2
+\sum_{k=1,2}\left((\partial_{\pm} r_k)^2 +r_k^2\,\eta_{k,2}^2\,
\partial_{\pm}\vec{\psi}^{\,\textrm{t}} M^{(k)}\,\partial_{\pm}\vec{\psi}\,\right)\right]\;,
\nonumber
\end{eqnarray}
up to $o(1/g)$ terms. Using the field redefinitions (\ref{t and Psi NFS}) and (\ref{phi sigma NFS})
with the choice (\ref{Pi=Omega}) for $\Pi$, now gives
\begin{equation}
\label{expansionT--U3}
T_{--}\,=\,\frac{1}{g}\left(
-\,(\partial_{-}\tau)^2+\,(\partial_{-}\chi)^2
+ \,(\partial_{-}\vec{z}\,)^2
+\sum_{k\,=1,2}\, (\partial_{-}\vec{r}_k)^2\right)+\,O\big(\,g^{-3/2}\,\big)\;.
\end{equation}
Notice that this result does not require to specify either $k_t$ or $\vec{k}_\phi$. Instaed, the vanishing of the term $O(g)$ in $T_{++}$ yields
\begin{equation}
\label{kt and kPsi}
k_t^2\,=\,k_\Psi^2\;.
\end{equation}
Letting $k_t= k_\Psi \equiv k$ and using the vector $\vec{k}_\phi$ solving (\ref{k vec U3}), we find
\begin{equation}
T_{++}=\,
-\,2\,k\,\partial_{+}(\,\tau\,-\,\chi\,)
-\,k^2
\left(z^2-\sum_{k=1,2} r_k^2\,\eta_{k,2}^2
\left( \vec{\mu}^{(k)\,\textrm{t}}_\omega \frac{\vec{k}_\phi}{k_\Psi} + m^{(k)}_\omega \right)\right)\;,
\end{equation}
up to terms infinitesimal in the limit $g\rightarrow \infty$. \\
Thus, with $k_t= k_\Psi \equiv k$, it becomes natural to introduce $\hat{\sigma}^+=k\,\sigma^+$, and the Polyakov action in the near-flat space limit reads
\begin{eqnarray}
\label{NFSlagrangianSigmaRescaled}
\lim_{g\,\rightarrow\,\infty}\,S  &=&
-\,2\int \left\{\,-\,\partial_{\hat{+}}\tau\,\partial_{-}\tau
+ \partial_{\hat{+}}\chi\,\partial_{-}\chi
+ \partial_{\hat{+}}\vec{z}\, \partial_{-}\vec{z}
+\sum_{i\,=\,1}^{2} \partial_{\hat+}\vec{r}_i\,\partial_{-}\vec{r}_i
\phantom{\frac{c^2p^2}{b^2}}\right.\\
\rule{0pt}{.76cm}
& &\hspace{2.4cm}
\left.-\,z^2\,\partial_{-}\tau\,-
\left[\;\sum_{k=1,2} r_k^2\,\eta_{k,2}^2
\left( \vec{\mu}^{(k)\,\textrm{t}}_\omega \frac{\vec{k}_\phi}{k_\Psi} + m^{(k)}_\omega \right)\,\right]
\partial_{-}\chi
\,\right\}\,
d\hat{\sigma}^+ d\sigma^-\,,\nonumber
\end{eqnarray}
while the Virasoro constraints $T_{--}=0$ and $T_{++}=0$, to the first non trivial order, give respectively
\begin{eqnarray}
& & 
-\,(\partial_{-}\tau)^2+\,(\partial_{-}\chi)^2
+ \,(\partial_{-}\vec{z}\,)^2
+\sum_{k\,=\,1}^{2} \,(\partial_{-}\vec{r}_k)^2\,=\,0\;,\\
\rule{0pt}{.8cm}& & 
2\,\partial_{\hat{+}}(\,\tau\,-\,\chi\,)
+\,
z^2-\sum_{k=1,2}
 r_k^2\,\eta_{k,2}^2
\left( \vec{\mu}^{(k)\,\textrm{t}}_\omega \frac{\vec{k}_\phi}{k_\Psi} + m^{(k)}_\omega \right)
\,=\,0\;.
\end{eqnarray}
In appendix \ref{appendix special cases}  we discuss in detail the application of the results obtained in this section for the relevant known cases.

\section*{Conclusions}

In this paper we studied the near-flat space limit for the bosonic sector of strings propagating in ten dimensional target spaces $AdS_5 \times M^5$ with different choices for the five dimensional internal manifold, like $T^{p,q}$, $Y^{p,q}$ and $L^{p,q,r}$. Since $p\sqrt[4]{\lambda}$ is kept fixed, this limit explores an intermediate region between the pp-wave and giant magnon regimes.

Our first result is that the bosonic sector of the limiting theory is the same found for $AdS_5 \times S^5$ in \cite{MaldacenaSwanson06}, at least for the stable Einstein spaces, which admit an unitary field theory dual. 
In addition, by introducing proper generalized metrics with $U(1)^3$ symmetry, we have shown that the coefficients characterizing the field redefinitions of the near-flat space limit are the same occurring in the coordinate transformations adopted to get the pp-wave metric as the Penrose limit of $AdS_5 \times M^5$.

We remark that the near-flat space limit of the fermionic sector for internal spaces different from $S^5$ remains to be studied. Indeed, the presence of the RR five form makes difficult to construct the explicit form of the IIB superstring action on $AdS_5 \times M^5$ in terms of the coordinate fields, even for the simplest case of $M^5=T^{1,1}$. However, our results strongly indicate that the near-flat space limit of the full string sigma model on $AdS_5 \times M^5$ with Sasaki-Einstein $M^5$s should be the same sigma model found for $AdS_5 \times S^5$ in \cite{MaldacenaSwanson06}. 

Our analysis could be generalized to study the near-flat space limit for the $\beta$-deformations of the backgrounds considered here.

The final aim of our work is to improve the understanding of the integrable structure underlying the AdS/CFT correspondence by identifying the features of the already known results that can be extended to the less supersymmetric cases. Much has still to be done in this direction.

\subsection*{Acknowledgments}

We would like to thank Francesco Bonsante, Niels Obers, Juan Maldacena, Konstantin Zarembo and in particular Augusto Sagnotti for stimulating discussions. We are grateful to Augusto Sagnotti also for useful comments on the draft. We thank the Galileo Galilei Institute for the kind hospitality and support during the last part of this work.\\
S.B. is supported in part by the National Science Foundation under Grant No. PHY-0243680.
E.T. is supported in part by a Fellowship from Scuola Normale Superiore, by INFN, by the MIUR-PRIN contract 2003-023852 and by the INTAS contract 03-51-6346.

\appendix
\section{The near-flat space limit of $AdS_5\times S^5$}
\label{S5}

In this appendix we briefly review the first part of the near-flat space limit analysis for type IIB string theory on $AdS_5 \times S^5$ performed by Maldacena and Swanson \cite{MaldacenaSwanson06}.
In order to compare this case with the ones presented in this paper, we consider only the bosonic sector.
The metric of $S^5$ can be written
\begin{equation}
\label{S5metric}
ds^2_{S^5}\,=\,R^2\Big(\cos^2\theta\;d\psi^2+d\theta^2+\sin^2\theta\;d\widetilde{\Omega}^2_3\,\Big)\;,
\end{equation}
where  $d\widetilde{\Omega}_3^2=d\gamma_1^2
+\cos^2 \gamma_1\,d\gamma_2^2+\sin^2 \gamma_1\,d\gamma_3^2$ is the metric of the unit three sphere.
To study the near-flat space limit of $AdS_5\times S^5$, they introduced the field redefinitions
\begin{eqnarray}
\label{field redefinitions S5}
& & t\,=\,\sqrt{g}\,\sigma^{+}+\,\frac{\tau}{\sqrt{g}}\;,
\hspace{2.5cm}
\rho\,=\,\frac{z}{\sqrt{g}}\;,\\
\rule{0pt}{.76cm}
& & \psi\,=\,\sqrt{g}\,\sigma^{+}+\,\frac{\chi}{\sqrt{g}}\;,
\hspace{2.4cm}
\theta\,=\,\frac{y}{\sqrt{g}}\;.
\nonumber
\end{eqnarray}
Taking the limit $g\rightarrow \infty$, one then finds that the leading term in the Lagrangian is proportional to $g\,(\partial_{-}\tau-\partial_{-}\chi)$, which however is a total derivative. Thus
\cite{MaldacenaSwanson06}
\begin{eqnarray}
\label{lagrangian finite part S5}
\lim_{g\,\rightarrow\,\infty}S 
&=&
-\,2\int \left\{\rule{0pt}{.4cm}
-\partial_{+}\tau\,\partial_{-}\tau
+ \partial_{+}\chi\,\partial_{-}\chi
+ \partial_{+}\vec{z}\, \partial_{-}\vec{z}
+ \partial_{+}\vec{y}\, \partial_{-}\vec{y}\right.\\
& & \hspace{8cm}\left.\rule{0pt}{.4cm}
-z^2\,\partial_{-}\tau
-y^2\,\partial_{-}\chi\,
\right\}\, d\sigma^+ d\sigma^- \,,
\nonumber
\end{eqnarray}
where 
\begin{eqnarray}
\label{zvec}
\partial_{+}\vec{z} \,\partial_{-}\vec{z}
&=&
\partial_{+}z\,\partial_{-}z+z^2
\big(\,\partial_{+}\delta_1\,\partial_{-}\delta_1
+\cos^2 \delta_1\,\partial_{+}\delta_2\,\partial_{-}\delta_2
+\sin^2 \delta_1\,\partial_{+}\delta_3\,\partial_{-}\delta_3\,\big)\;,
\\
\partial_{+}\vec{y} \,\partial_{-}\vec{y} 
&=&
\partial_{+}y\,\partial_{-}y+y^2
\big(\,\partial_{+}\gamma_1\,\partial_{-}\gamma_1
+\cos^2 \gamma_1\,\partial_{+}\gamma_2\,\partial_{-}\gamma_2
+\sin^2 \gamma_1\,\partial_{+}\gamma_3\,\partial_{-}\gamma_3\,\big)\;.
\phantom{xxxx}
\end{eqnarray}
This action is right conformally invariant, i.e. it is invariant under $\sigma^- \rightarrow f(\sigma^-)$, but is not invariant under left conformal transformations.
As for the Virasoro constraints $T_{--}=0$ and $T_{++}=0$ (see (\ref{Virasoro})), imposing the vanishing of the first non trivial term of their expansion at large $g$, one finds the two equations
\begin{eqnarray}
\label{S5T--Final}
& & 
-\,(\partial_{-}\tau)^2+\,(\partial_{-}\chi)^2
+ \,(\partial_{-}\vec{z}\,)^2
+\, (\partial_{-}\vec{y}\,)^2
\,=\,0\;,
\\
\label{S5T++Final}
\rule{0pt}{.7cm}& & 
2
\,\partial_{+}(\,\tau\,-\,\chi\,)
+\,z^2+\,y^2\,=\,0\;.
\end{eqnarray}
Using these conditions and suitable worldsheet coordinates, one finally arrives at a gauge fixed Lagrangian which has been employed to study the $S$ matrix at one \cite{KloseZarembo07} and two loops \cite{KloseMinahanZarembo07}.

\section{Special cases}
\label{appendix special cases}

In this appendix we recover the Penrose and near-flat space limits for $T^{p,q}$ and $Y^{p,q}$ as special cases of generalized $U(1)^3$ metrics, applying the results obtained in sections \ref{section U3}. In the last subsection we explicitly consider the special case of
the $L^{p,q,r}$ metrics.

\subsection{The $T^{p,q}$ case}
\label{appendix Tpq case}

The metrics (\ref{abcmetric}) in the coordinates $(\psi,\phi_1,\phi_2,\theta_1,\theta_2)$ have strictly positive $g_{44} (\vec{\theta}_0)$ and $g_{55} (\vec{\theta}_0)$, therefore we cannot directly apply the results of section \ref{section U3}, but only after the change of coordinates $y_i = \cos \theta_i$ for $i=1,2$. In this subsection we apply the procedure explained in section \ref{section U3} to a metric with $U(1)^3$ symmetry satisfying conditions which are slightly different  with respect to those introduced in section \ref{section U3}. This metric allows to recover the results of section \ref{NFS for Tpq} as a special case in a direct way.\\
In particular, given a null geodesic in $AdS_5\times M^5$ having $\rho = 0$ and $\vec{\theta}^{\textrm{t}}=\vec{\theta}_0^{\textrm{t}}=(\theta_{1,0},\theta_{2,0})$, now we take (\ref{metricsigma}) with $g_{44} (\vec{\theta}_0)>0$ and $g_{55} (\vec{\theta}_0)>0$. The expansion of the ten dimensional metric is then
\begin{eqnarray}
\label{ds2serieTpqSigma}
ds^2 &  &\hspace{-.5cm}=\, R^2\left[\,-\,dt^2+g_{ab} (\vec{\theta}_0)\,d\psi_a \,d\psi_b\,\right]
-z^2 dt^2 +d\vec{z}^{\,2} 
+R^2 g_{44} (\vec{\theta}_0)\,d\theta_1^2
+ R^2 g_{55} (\vec{\theta}_0)\,d\theta_2^2
\\
& &\rule{0pt}{.7cm}
+R^2\,\partial_k \,g_{ab} (\vec{\theta}_0)\,(\theta_k-\theta_{k,0})\,d\psi_a \,d\psi_b
+R^2\,\frac{\partial_k\partial_p \,g_{ab} (\vec{\theta}_0)}{2}\,
(\theta_k-\theta_{k,0})(\theta_p-\theta_{p,0})\,d\psi_a \,d\psi_b\;.
\nonumber 
\end{eqnarray}
Then, we also assume
\begin{equation}
\label{condition1Tpq}
\partial_k \,g_{ab}(\vec{\theta}_0)
=\,0
\hspace{2cm}k\,=\,1,2
\end{equation}
which induce the following definitions of the coordinates $r_k$
\begin{equation}
\label{rhozeta}
\theta_k-\theta_{k,0}\,=\,\eta_{k,1}\,\frac{r_k}{R}
\hspace{1.7cm}
k\,=\,1,2
\end{equation}
where $\eta_{k,1}$ are constants. Another assumption we make is
\begin{equation}
\label{condition2Tpq}
\partial_1 \partial_2\, g_{ab}(\vec{\theta}_0)
=\,0\;,
\end{equation}
in order to avoid a term containing $r_1 r_2$ after the limit $R\rightarrow\infty$.
The terms $d\theta_k^2$ in (\ref{ds2serieTpqSigma}) suggest that the most  convenient choice for  $\eta_{k,1}$ is
\begin{equation}
\label{etaTpq}
\eta_{1,1}\,=\,g_{44} (\vec{\theta}_0)^{-\frac{1}{2}}\;,
\hspace{2cm}
\eta_{2,1}\,=\,g_{55} (\vec{\theta}_0)^{-\frac{1}{2}}\;.
\end{equation}
At this point the expansion of the metric of $AdS_5\times M^5$ for large $R$ becomes
\begin{equation}
\label{ds2penroseTpqSigma0}
ds^2 =\, R^2\left[\,-\,dt^2+g_{ab} (\vec{\theta}_0)\,d\psi_a \,d\psi_b\,\right]
-z^2 dt^2 +d\vec{z}^{\,2}
+\sum_{k=1,2}\left(dr^2_k+\,\frac{\eta_{k,1}^2}{2}\,r_k^2\;
d\vec{\psi}^{\textrm{t}} N^{(k)}d\vec{\psi}\,\right)\,,
\end{equation}
where
\begin{equation}
\label{N(k)}
N^{(k)}\,\equiv\,
\partial_k^2 \,g_{ab}(\vec{\theta}_0)
\hspace{2cm}k\,=\,1,2\;.
\end{equation}
Comparing (\ref{ds2penroseTpqSigma0}) with (\ref{ds2penroseU30}), it becomes clear that hereafter the procedure is exactly the same as in section \ref{section penrose U3} but with 
$N^{(k)}$ instead of $M^{(k)}$ and with $\eta^2_{k,1}/2$ instead of  $\eta^2_{k,2}$. \\

\noindent As for the near-flat space limit for this $U(1)^3$ metric, the fields $r_k$ are now defined as
\begin{equation}
\label{rhozetaNFSTpqsigma}
\theta_k-\theta_{k,0}\,=\,\eta_{k,1}\,\frac{r_k}{\sqrt{g}}
\hspace{1.7cm}
k\,=\,1,2
\end{equation}
with $\eta_{k,1}$ given by (\ref{etaTpq}), while all the other redefinitions are the usual ones. The expansion of the Lagrangian now reads
\begin{eqnarray}
\label{LagrangianTpqsigma1}
\mathcal{L} & = &
-\,2\,\left\{\,g\,\big[-\partial_{+} t \,\partial_{-} t +\,\partial_{+} \Psi \,\partial_{-} \Psi\,\big]
\phantom{\frac{\eta_{k,1}^2}{2}}\right.
\\
\rule{0pt}{.5cm}& & \hspace{1.1cm}
\left.-\,z^2 \,\partial_{+} t \,\partial_{-} t  + \partial_{+} \vec{z} \,\partial_{-} \vec{z} 
+\sum_{k=1,2}\left(\partial_{+} r_k \,\partial_{-} r_k +r_k^2\;\frac{\eta_{k,1}^2}{2}\;
\partial_{-}\vec{\psi}^{\,\textrm{t}} N^{(k)}\,\partial_{+}\vec{\psi}\,\right)\right\}\;,
\nonumber
\end{eqnarray}
where $N^{(k)}$ are the matrices (\ref{N(k)}). Likewise for the Penrose limit, also for the near-flat space limit we can apply the formulas obtained in section \ref{section NFS U3}, provided that one uses $N^{(k)}$ instead of $M^{(k)}$ and $\eta^2_{k,1}/2$ instead of  $\eta^2_{k,2}$.\\

\noindent {\bf The Penrose limit of the $T^{p,q}$ metrics.} Here, specializing our discussion to (\ref{abcmetric}), we recover the coordinate transformations found in \cite{KlebanovItzhaki02}.
The null geodesic for the ten dimensional metric is described by $\rho=0$, $\theta_1=\theta_2=0$ and $t=\Psi$ with 
$\Psi=a(\psi+p\,\phi_1+q\,\phi_2)$. Having checked that the metrics (\ref{abcmetric}) satisfy all the assumptions made throughout the above discussion, eqs. (\ref{lambda vec U3}) with the proper substitutions ($N^{(k)}$ instead of $M^{(k)}$ and of $\eta^2_{k,1}/2$ instead of  $\eta^2_{k,2}$) become respectively
\begin{equation}
\label{lambdavecTpq}
\left(\begin{array}{cc}
2\,b^2 & 0\\
0 & 0
\end{array}\right)\vec{\lambda}\,=\,
\left(\begin{array}{c}
a\,p \\
0
\end{array}\right)\;,
\hspace{1.5cm}
\left(\begin{array}{cc}
0 & 0\\
0 & 2\,c^2
\end{array}\right)\vec{\lambda}\,=\,
\left(\begin{array}{c}
0 \\
a\,q
\end{array}\right)\;.
\end{equation}
Their solution
\begin{equation}
\label{lambda solution Tpq}
\lambda_1\,=\,\frac{a\,p}{2\,b^2}\;,
\hspace{2cm}
\lambda_2\,=\,\frac{a\,q}{2\,c^2}\;,
\end{equation}
gives the vector to use in (\ref{def lambda and Omega}).
Notice that $\textrm{det}\,H^{(k)}_\omega=0$ for $k=1,2$, as expected. Given 
(\ref{lambda solution Tpq}), one finds that
\begin{equation}
\frac{\eta_{1,1}^2}{2}
\left(\,\vec{\mu}^{(1)\,\textrm{t}}_\omega \vec{\lambda} + m^{(1)}_\omega\right)
\,=\, -\,\frac{a^2 p^2}{4\,b^4}\;,
\hspace{1.3cm}
\frac{\eta_{2,1}^2}{2}
\left(\,\vec{\mu}^{(2)\,\textrm{t}}_\omega \vec{\lambda} + m^{(2)}_\omega\right)
\,=\, -\,\frac{a^2 q^2}{4\,c^4}
\;.
\end{equation}
Finally, comparing the matrices $H^{(k)}_\omega$ that can be read from (\ref{lambdavecTpq}) with (\ref{OmegaU3}) properly adapted to the $U(1)^3$ metric we are considering, one can easily conclude that the $\Omega$ to adopt in (\ref{def lambda and Omega}) for this case is the identity matrix. \\
At this point it is straightforward also to specialize the formulas of subsection \ref{section NFS U3} for (\ref{abcmetric}) and recover the results of section \ref{NFS for Tpq} for the near-flat space limit of 
$T^{p,q}$.

\subsection{The $Y^{p,q}$ case}
\label{appendix Ypq case}

When $M^5$ is a $Y^{p,q}$ manifold with metric (\ref{Ypqmetric}), the null geodesic in the ten dimensional space is given by $\rho=0$, $\theta=0$ and $y=y_0$ such that $p(y_0)=0$. Thus, the $Y^{p,q}$ metrics (\ref{Ypqmetric}) written in the usual coordinates $(\psi,\phi,\beta,\theta, y)$ are not included either in the generalized $U(1)^3$ metric considered in section \ref{section penrose U3} or in the one introduced in the subsection \ref{appendix Tpq case}, but they fall between them. Therefore, we can introduce a $U(1)^3$ metric satisfying mixed assumptions, namely with $g_{44} (\vec{\theta}_0)>0$ and $\tilde{g}_{55} (\vec{\theta}_0)=\partial_1 \,\tilde{g}_{55}(\vec{\theta}_0)=0$ but $\partial_2 \,\tilde{g}_{55}(\vec{\theta}_0) \neq\,0$, and we also assume that
\begin{equation}
\label{condition1Ypq}
\partial_1 \,g_{ab}(\vec{\theta}_0) =\,0\;,
\hspace{2cm}
\partial_2 \,g_{ab}(\vec{\theta}_0) \neq\,0\;.
\end{equation}
Thus, as we have learned from the previous case, we introduce the coordinates $r_k$ via
\begin{equation}
\label{rhozetaYpq}
\theta_1-\theta_{1,0}\,=\,\eta_{1,1}\,\frac{r_1}{R}\;,
\hspace{1.7cm}
\theta_2-\theta_{2,0}\,=\,\eta_{2,2}\,\frac{r_2^2}{R^2}
\end{equation}
where the constants $\eta_{1,1}$ and $\eta_{2,2}$ are defined in (\ref{etaTpq}) and (\ref{etaU3}), respectively. The procedure to analyze the Penrose limit and the near-flat space limit in this case is the same of section \ref{section U3} but now only the matrix $M^{(1)}$ must be replaced by the matrix $N^{(1)}$ defined in (\ref{N(k)}) and only $\eta_{1,2}^2$ must be replaced by $\eta_{1,1}^2/2$.\\

\noindent {\bf The Penrose limit for the $Y^{p,q}$ metrics.} Specifying the analysis of the Penrose limit to the $Y^{p,q}$ metrics (\ref{Ypqmetric}),
we have $\vec{\theta}^{\textrm{t}}=(\theta,y)$ and $\vec{\psi}^{\textrm{t}}=(\psi,\phi,\beta)$.
The null geodesic for the ten dimensional metric is given by $\rho=0$, $\theta=0$, $y=y_0$ such that $p(y_0)=0$ and $t=\Psi$ with 
$\Psi=(\psi-(1-c\,y_0)\phi+ y_0\,\beta)/3$. The metric (\ref{Ypqmetric}) verifies all the assumptions made throughout the above discussion, and therefore we can apply the final expressions.\\
In particular, $\tilde{g}_{55}(\theta_1,\theta_2)=6\,p(y)$ depends only on $\theta_2$, and the condition
$\partial_2 \,\tilde{g}_{55}(\vec{\theta}_0) \neq\,0$ becomes $p'(y_0)\neq 0$. The constants $\eta_{1,1}$ and $\eta_{2,2}$ read
\begin{equation}
\label{nuYpq}
\eta_{1,1}\,=\,\left(\frac{1-c\,y_0}{6}\right)^{-\frac{1}{2}}\;,
\hspace{2cm}
\eta_{2,2}\,=\,\frac{3}{2}\;p'(y_0)\;.
\end{equation}
The eqs. (\ref{lambda vec U3}), once adapted to the $U(1)^3$ metric we are considering, become respectively
\begin{eqnarray}
\label{lambdavecYpq1}
\frac{1-c\,y_0}{3}
\left(\begin{array}{cc}
1 & 0\\
0 & 0
\end{array}\right)\vec{\lambda}\,+
\,\frac{1-c\,y_0}{3}
\left(\begin{array}{c}
1 \\
0
\end{array}\right)
&=&
\left(\begin{array}{c}
0 \\
0
\end{array}\right)\;,
\\
\label{lambdavecYpq2}
\rule{0pt}{1cm}
\frac{p'(y_0)}{6}
\left(\begin{array}{cc}
c & c^2\\
1 & c
\end{array}\right)\vec{\lambda}
+\frac{1}{3}
\left(\begin{array}{c}
c \\
1
\end{array}\right)
&=&
\left(\begin{array}{c}
0 \\
0
\end{array}\right)\;,
\end{eqnarray}
and their solution is
\begin{equation}
\label{lambdasolutionYpq}
\lambda_1\,=\,-\,1\;,
\hspace{2cm}
\lambda_2\,=\,c-\frac{2}{p'(y_0)}\;.
\end{equation}
Reading $H^{(k)}$ from (\ref{lambdavecYpq1}) and (\ref{lambdavecYpq2}), one can verify that $\textrm{det}\,H^{(k)}=0$ for $k=1,2$, as expected. With $\vec{\lambda}$ given by (\ref{lambdasolutionYpq}), we get
\begin{equation}
\label{YpqPenrose muterm}
\frac{1}{2\,g_{44} (\vec{\theta}_0)}
\left(\,\vec{\mu}^{(1)\,\textrm{t}}_\omega \vec{\lambda} + m^{(1)}_\omega\right)
\,=\,
\frac{\partial_2 \tilde{g}_{55} (\vec{\theta}_0)}{4}
\left(\,\vec{\mu}^{(2)\,\textrm{t}}_\omega \vec{\lambda} + m^{(2)}_\omega\right)
\,=\,-\,1
\;.
\end{equation}
Finally, the matrix $\Omega$ for the $Y^{p,q}$ satisfying the properly adapted version of (\ref{OmegaU3}) is
\begin{equation}
\Omega\,=\,
\left(\begin{array}{cc}
-1 & 0\\
c & -2/p'(y_0)
\end{array}\right)\;.
\end{equation}
 Summarizing, aside from $\rho=z/R$, the change of coordinates which allows to find the pp-wave metric (\ref{ppwavemetric}) as large $R$ limit of  the metrics (\ref{Ypqmetric}) is
\begin{eqnarray}
\label{field redefinitions Ypq penrose}
& & \psi\,=\,\left(2+\frac{2\,y_0}{p'(y_0)}\right)\Psi\,
-\varphi_1+\,\frac{2\,y_0}{p'(y_0)}\, \varphi_2\;,\\
\rule{0pt}{.9cm}
& & 
\phi\,=\,-\,\Psi\,- \,\varphi_1\;,
\hspace{6.4cm}
\theta\,=\,\left(\frac{6}{1-c\,y_0}\right)^{\frac{1}{2}}\frac{r_1}{R}\;,
\nonumber\\
\rule{0pt}{.9cm}
& & 
\beta\,=\,\left(c-\frac{2}{p'(y_0)}\right)\Psi\,
+c\,\varphi_1\,-\,\frac{2}{p'(y_0)}\,\varphi_2\;,
\hspace{1.8cm}\,
y\,=\,y_0+\frac{3}{2}\,p'(y_0)\,\frac{r_2^2}{R^2}\;,
\nonumber
\end{eqnarray}
where $\Psi$ and $t$ are given in (\ref{x^pm}).\\ 
Let us remark that that in finding the pp-wave metric (\ref{ppwavemetric}) we have not made use of the explicit expression of $p(y)$ for $Y^{p,q}$, but just of the conditions $p(y_0)=0$ and $p'(y_0)\neq 0$.
Imposing $p'(y_0)=-\,2\,y_0$ into (\ref{field redefinitions Ypq penrose}), we recover the change of coordinates found in \cite{KupersteinSonnenschein06}.\\
Even for the Penrose limit, one can recover the pp-wave changes of coordinates for $T^{1,1}$ as a special case of (\ref{field redefinitions Ypq penrose}) letting $c=0$, $p'(y_0)=-\,2\,y_0$ and taking into account that
$\theta=\theta_1$, $\phi=-\,\phi_1$, $y=\cos\theta_2$ and $\beta=\phi_2$, as was done for the near-flat space in the last part of section \ref{NFS Ypq}.

\subsection{The $L^{p,q,r}$ metrics}
\label{Lpqrapp}

In this appendix we study the near-flat space limit of the Polyakov action and of the Virasoro constraints for $AdS_5 \times L^{p,q,r}$, treating these target space metrics as special cases of the generalized $U(1)^3$ metric introduced in subsection \ref{appendix Ypq case}.\\
The $L^{p,q,r}$ metrics in the canonical form \cite{CveticLuPagePope05} are
\begin{eqnarray}
\label{Lpqrmetric}
ds^2_{M^5} &=& R^2
\left\{\,\left[\,d\psi+\,\frac{\alpha-x}{\alpha}\,\sin^2\theta\,d\phi+
\,\frac{\beta-x}{\beta}\,\cos^2\theta\,d\gamma\,\right]^2
+\frac{\Delta_\theta-x}{\Delta_\theta}\,d\theta^2 +\frac{\Delta_\theta-x}{4\,\Delta_x}\,dx^2
\right.\nonumber\\
\rule{0pt}{.8cm}& & \hspace{-1.9cm}
\left.+\,\frac{\Delta_x}{\Delta_\theta-x}
\left(\,\frac{\sin^2\theta}{\alpha}\,d\phi+\,\frac{\cos^2\theta}{\beta}\,d\gamma\right)^2
+\,\frac{\Delta_\theta}{\Delta_\theta-x}\,\cos^2\theta\,\sin^2\theta
\left(\,\frac{\alpha-x}{\alpha}\,d\phi-\,\frac{\beta-x}{\beta}\,d\gamma\right)^2
\,\right\},\nonumber\\
& & 
\end{eqnarray}
where $\Delta_\theta = \alpha\,\cos^2\theta + \beta\,\sin^2\theta$, but we keep $\Delta_x$ as a generic function of $x$ for our purposes.
For the coordinate $\theta$ we have $0\leqslant \theta \leqslant \pi/2$, while $x$ lies in the interval $[x_1,x_2]$, whose endpoints are two adjacent real roots of $\Delta_x$ and $\Delta_x \geqslant 0$. We can require $x_1\geqslant 0$ and also that $\alpha > x_2$, $\beta \geqslant x_2$.\\
The null geodesic we consider is characterized by $\rho=0$, $\theta=0$ and $x=x_0$ such that $\Delta_x|_{x_0}=0$ (i.e. $x_0$ is either $x_1$ or $x_2$) and $t=\Psi$ with $\Psi=\psi+(1-x_0/\beta)\gamma$.\\
The $L^{p,q,r}$ metrics (\ref{Lpqrmetric}) are special cases of the generalized metrics introduced in subsection \ref{appendix Ypq case}. In particular, $\vec{\theta}^{\textrm{t}}=(\theta,x)$, $\vec{\psi}^{\textrm{t}}=(\psi,\phi,\gamma)$ and $\tilde{g}_{55}(\theta_1,\theta_2)=4\,\Delta_x/(\Delta_\theta-x)$ depends on both $\theta_1$ and $\theta_2$, but the condition $\partial_2 \,\tilde{g}_{55}(\vec{\theta}_0) \neq\,0$ becomes $\Delta'_{0}\equiv \Delta'_{x}|_{x_0}\neq 0$.\\

\noindent {\bf The Penrose limit of the $L^{p,q,r}$ metrics.} To study the Penrose limit, we repeat the procedure explained in subsection \ref{appendix Ypq case} for (\ref{Lpqrmetric}). Using the notation of \cite{KupersteinSonnenschein06}, where the constants $a_0$, $b_0$ and $c_0$ are defined as
\begin{equation}
a_0\,=\,\frac{\alpha(\beta-x_0)}{\Delta'_{0}}\;,
\hspace{.4cm}
b_0\,=\,\frac{\beta(\alpha-x_0)}{\Delta'_{0}}\;,
\hspace{.4cm}
c_0\,=\,-\,\frac{(\beta-x_0)(\alpha-x_0)}{\Delta'_{0}}\;,
\hspace{.4cm}
\textrm{with}
\hspace{.4cm}
\Delta'_{0}\,\equiv\,\Delta'_x|_{x_0}
\end{equation}
one finds
\begin{equation}
\label{nuLpqr}
\eta_{1,1}\,=\,\left(\frac{\alpha-x_0}{\alpha}\right)^{-\frac{1}{2}},
\hspace{2cm}
\eta_{2,2}\,=\,\frac{\Delta'_{0}}{\alpha-x_0}\;.
\end{equation}
In this case, the properly adapted version of eqs. (\ref{lambda vec U3}) become respectively
\begin{eqnarray}
\label{lambdavecLpqr1}
2\left(\begin{array}{cc}
(\alpha-x_0)/\alpha & (x_0-\beta)/\beta\\
(x_0-\beta)/\beta & [\alpha(\beta-x_0)^2]/[\beta^2(\alpha-x_0)]
\end{array}\right)\vec{\lambda}\,+\,2
\left(\begin{array}{c}
(\alpha-x_0)/\alpha \\
(x_0-\beta)/\beta
\end{array}\right)
&=&
\left(\begin{array}{c}
0 \\
0
\end{array}\right)
\phantom{xxxx}
\\
\label{lambdavecLpqr2}
\rule{0pt}{1cm}
\frac{1}{\beta}
\left(\begin{array}{cc}
0 & 0\\
0 & 1/b_0
\end{array}\right)\vec{\lambda}
-\,\frac{1}{\beta}
\left(\begin{array}{c}
0 \\
1
\end{array}\right)
&=&
\left(\begin{array}{c}
0 \\
0
\end{array}\right)
\end{eqnarray}
and their solution is
\begin{equation}
\label{lambdaYpqSigma}
\lambda_1\,=\,-\,1+\,a_0\;,
\hspace{2cm}
\lambda_2\,=\,b_0\;.
\end{equation}
Now one can check that $\textrm{det}\,H^{(k)}=0$ for $k=1,2$ and that, given the solution (\ref{lambdaYpqSigma}) for $\vec{\lambda}$, the two equation in (\ref{YpqPenrose muterm}) hold 
also in this case.\\
The matrix $\Omega$ solving the properly adapted version of (\ref{OmegaU3}) reads
\begin{equation}
\Omega\,=\,
\left(\begin{array}{cc}
a_0 & 1\\
b_0 & 0
\end{array}\right)\;.
\end{equation}
Summarizing, aside from $\rho=z/R$, the change of coordinates in the metric of $AdS_5 \times L^{p,q,r}$ giving the pp-wave metric (\ref{ppwavemetric}) when $R \rightarrow \infty$
is \footnote{\,Here we correct a misprint in eq. (4.3) of \cite{KupersteinSonnenschein06}.}
\begin{eqnarray}
\label{field redefinitions Lpqr penrose}
& & \psi\,=\,\big(1+c_0\big)\Psi\,+c_0\,\varphi_1\;,\\
\rule{0pt}{1cm}
& & 
\phi\,=\,\big(-1+a_0\big)\Psi\,+\,a_0\,\varphi_1+\,\varphi_2\;,
\hspace{2.1cm}
\theta\,=\,\left(\frac{\alpha}{\alpha- x_0}\right)^{\frac{1}{2}}\frac{r_1}{R}\;,
\nonumber\\
\rule{0pt}{.8cm}
& & 
\gamma\,=\,b_0\,\Psi\,
+\,b_0\,\varphi_1\;,
\hspace{4.7cm}\,
x\,=\,x_0+\frac{\Delta'_0}{\alpha-x_0}\;\frac{r_2^2}{R^2}\;,
\nonumber
\end{eqnarray}
where $\Psi$ and $t$ are given in (\ref{x^pm}).
We remark that we have not made use of the explicit expression of $\Delta_x$ for $L^{p,q,r}$, but only of the conditions $\Delta_{x}|_{x_0}=0$ and $\Delta_{x}'|_{x_0}\neq 0$.\\

\noindent {\bf The near-flat space limit of the $L^{p,q,r}$ metrics.} As shown in section \ref{section NFS U3}, the coefficients characterizing the field redefinitions of the near-flat space limit are the same occurring in the change of coordinates leading to the pp-wave metric when $R \rightarrow \infty$. In particular, the field redefinitions involved in the near-flat space limit for $AdS_5 \times L^{p,q,r}$ are
\begin{eqnarray}
\label{field redefinitions Lpqr}
& & t\,=\,k_t \sqrt{g}\,\sigma^{+}+\,\frac{\tau}{\sqrt{g}}\;,
\hspace{5.2cm}
\rho\,=\,\frac{z}{\sqrt{g}}\;,\nonumber\\
\rule{0pt}{.76cm}
& & \psi\,=\,k_\psi \sqrt{g}\,\sigma^{+}+\,\frac{K_\psi\,\chi}{K\sqrt{g}}\,+c_0 \,\varphi_1\;,\\
\rule{0pt}{.76cm}
& & 
\phi\,=\,k_{\phi} \sqrt{g}\,\sigma^{+}+\,\frac{K_{\phi}\,\chi}{K\sqrt{g}}\,+a_0\,\varphi_1+\varphi_2\;,
\hspace{2.2cm}
\theta\,=\,\left(\frac{\alpha}{\alpha-x_0}\right)^{\frac{1}{2}}\frac{r_1}{\sqrt{g}}\;,
\nonumber\\
\rule{0pt}{.87cm}
& & 
\gamma\,=\,k_{\gamma} \sqrt{g}\,\sigma^{+}+\,\frac{K_{\beta}\,\chi}{K\sqrt{g}}\,
+b_0\,\varphi_1\;,
\hspace{3.1cm}\,
x\,=\,x_0+\frac{\Delta'_{0}}{\alpha-x_0}\;\frac{r_2^2}{g}\;,
\nonumber
\end{eqnarray}
where $K=K_\psi+(1-x_0/\beta)K_\gamma$. 
The finiteness of the Polyakov action in the limit $g\rightarrow \infty$ provides two equations, whose solution is
\begin{equation}
\label{kvaluesLpqr}
k_\psi\,=\,\left(1-\,\frac{\beta-x_0}{\beta}\,b_0\right)k_\Psi\;,
\hspace{.7cm}
k_\phi\,=\,\big(-1+a_0\,\big)\,k_\Psi\;,
\hspace{1.2cm}
k_\gamma\,=\,b_0\,k_\Psi\;,
\end{equation}
where
\begin{equation}
\label{kPsiLpqr}
k_\Psi\,=\,k_\psi + \,\frac{\beta-x_0}{\beta}\,k_\gamma
\end{equation}
is a free parameter.
These expressions lead to the form (\ref{NFSYpqHprime}) for the Polyakov action.\\
Concerning the Virasoro constraints, the expansion of
$T_{--}$ at large $g$ is given by (\ref{expansionT--U3}), just using (\ref{field redefinitions Lpqr}), as expected from the general discussion of section \ref{section NFS U3}. Instead, choosing $k_t= k_\Psi$, the expansion of $T_{++}$ becomes (\ref{T++Ypq}) also for the $L^{p,q,r}$ metrics.\\
Thus, also these results for the near-flat space limit have been obtained without making use of the explicit expression for $\Delta_x$, but only assuming that $\Delta_x|_{x_0}=0$ and $\Delta'_x|_{x_0}\neq 0$. For completeness, the function $\Delta_x$ which makes (\ref{Lpqrmetric}) an Einstein manifold (i.e. with $R_{ab}=(4/R^2)g_{ab}$)  is \cite{CveticLuPagePope05}
\begin{equation}
\Delta_x\,=\,x(\alpha-x)(\beta-x)-\mu\;,
\end{equation}
where $\mu$ is a parameter that can be set to any nonzero value by rescaling $x$, $\alpha$ and $\beta$. The round sphere $S^5$ corresponds to $\mu=0$.


\begin{thebibliography}{99}




\bibitem{BMN02}
D. Berenstein, J. M. Maldacena and H. Nastase, 
``Strings in flat space and pp-wave from $\mathcal{N}=4$ super Yang-Mills'',  
JHEP {\bf 04} (2002) 013, [hep-th/0202021].


\bibitem{GubserKlebanovPolyakov02}
S. Gubser, I. Klebanov and A. Polyakov, 
``A semiclassical limit of the gauge/string correspondence'',
Nucl. Phys. B {\bf 636} (2002) 99, [hep-th/0204051].


\bibitem{MinahanZarembo}
J. Minahan and K. Zarembo,
``The Bethe ansatz for $\mathcal{N}=4$ super Yang-Mills'',  
JHEP {\bf 03} (2003) 013, [hep-th/0212208].


\bibitem{BenaRoibanPolchinski02}
I. Bena, J. Polchinski and R. Roiban,
``Hidden symmetries of $AdS_5\times S^5$\,'',  
Phys. Rev. D {\bf 69} (2004) 046002, [hep-th/0305116].


\bibitem{BeisertStaudacher03}
N. Beisert and M. Staudacher,
``The $\mathcal{N}=4$ SYM integrable super spin chain'',  
Nucl. Phys. B {\bf 670} (2003) 439, [hep-th/0307042].


\bibitem{Kruczenski04}
M. Kruczenski,
``Spin chains and string theory'',  
Phys. Rev. Lett. {\bf 93} (2004) 161602, [hep-th/0311203].


\bibitem{BeisertDippelStaudacher04}
N. Beisert, V. Dippel and M. Staudacher,
``A novel long range spin chain and planar $\mathcal{N}=4$ SYM super Yang Mills'',  
JHEP {\bf 07} (2004) 075, [hep-th/0405001].


\bibitem{ArutyunovFrolovStaudacher04}
G Arutyunov, S. Frolov and M. Staudacher
``Bethe ansatz for quantum strings'',  
JHEP {\bf 10} (2004) 016, [hep-th/0406256].


\bibitem{Beisert04review}
N. Beisert,
``The dilatation operator of $\mathcal{N}=4$ SYM super Yang Mills theory and integrability'',  
Phys. Rept.  {\bf 405} (2005) 1, [hep-th/0407277].


\bibitem{Beisert05 SU(2|2)}
N. Beisert,
``The $su(2|2)$ dynamic $S$ matrix'',  
[hep-th/0511082].


\bibitem{Janik:2006dc}
  R.~A.~Janik,
  ``The AdS(5) x S**5 superstring worldsheet S-matrix and crossing symmetry,''
  Phys.\ Rev.\  D {\bf 73}, 086006 (2006)
  [arXiv:hep-th/0603038].


\bibitem{Hernandez:2006tk}
  R.~Hernandez and E.~Lopez,
  ``Quantum corrections to the string Bethe ansatz,''
  JHEP {\bf 0607}, 004 (2006)
  [arXiv:hep-th/0603204].


\bibitem{HofmanMaldacena06}
D. Hofman and J. Maldacena,
``Giant magnons'',  
J. Phys. A {\bf 39} (2006) 13095, [hep-th/0604135].


\bibitem{BeisertHernandezLopez06}
N. Beisert, R. Hernandez and E. Lopez,
``A crossing symmetric phase for $AdS_5\times S^5$ strings'',  
JHEP {\bf 11} (2006) 070, [hep-th/0609044].

\bibitem{BeisertEdenStaudacher06}
N. Beisert, B. Eden and M. Staudacher,
``Transcendentality and crossing'',  
J. Stat. Mech. {\bf 01} (2007) P021, [hep-th/0610251].



\bibitem{MaldacenaSwanson06}
J. Maldacena and I. Swanson, 
``Connecting giant magnons to pp-wave: an interpolating limit  of $AdS_5\times S^5$\,'', 
[hep-th/0612079].


\bibitem{KloseZarembo07}
T. Klose and K. Zarembo,
``Reduced sigma model on $AdS_5\times S^5$: one-loop scattering amplitudes'',  
JHEP {\bf 02} (2007) 071, [hep-th/0701240].


\bibitem{KloseMinahanZarembo07}
T. Klose, T. McLoughlin, J. Minahan and K. Zarembo,
``Worldsheet scattering in $AdS_5\times S^5$ at two loops'',  
[arXiv:0704.3891].


\bibitem{KlebanovWitten98}
I. Klebanov and E. Witten,
``Superconformal field theory on threebranes at a Calabi-Yau singularity'',
Nucl. Phys.  B {\bf 536} (1998) 199, [hep-th/9807080].



\bibitem{GauntlettMartelli}
J. P. Gauntlett, D. Martelli, J. Sparks and D. Waldram,   
``Sasaki-Einstein metrics on $S^2\times S^3$\,'', 
Adv. Theor. Math. Phys. {\bf 8} (2004) 711, [hep-th/0403002].


\bibitem{CveticLuPagePope05}
M. Cvetic, H. Lu, D.N. Page and C.N. Pope
``New Einstein-Sasaki spaces in five and higher dimensions'', 
Phys. Rev. Lett. {\bf 95} (2005) 071101, [hep-th/0504225].


\bibitem{Martelli:2005wy}
  D.~Martelli and J.~Sparks,
  ``Toric Sasaki-Einstein metrics on S**2 x S**3,''
  Phys.\ Lett.\  B {\bf 621}, 208 (2005)
  [arXiv:hep-th/0505027].


\bibitem{BenvenutiHananyMartelli04}
S. Benvenuti, S. Franco, A. Hanany, D. Martelli and J. Sparks,
``An infinite family of superconformal quiver gauge theories with Sasaki-Einstein duals'', 
JHEP {\bf 06} (2005) 064, [hep-th/0411264].



\bibitem{BenvenutiKruczenskiLpqr}
S. Benvenuti and M. Kruczenski,
``From Sasaki-Einstein spaces to quivers via BPS geodesics: $L^{p,q|r}\,$'', 
JHEP {\bf 04} (2006) 033, [hep-th/0505206].


\bibitem{Franco:2005sm}
  S.~Franco, A.~Hanany, D.~Martelli, J.~Sparks, D.~Vegh and B.~Wecht,
  ``Gauge theories from toric geometry and brane tilings,''
  JHEP {\bf 0601}, 128 (2006)
  [arXiv:hep-th/0505211].


\bibitem{ButtiForcellaZaffaroniLpqr}
A. Butti, D. Forcella and A. Zaffaroni,
``The dual superconformal theory to $L^{p,q,r}$ manifolds'', 
JHEP {\bf 09} (2005) 018, [hep-th/0505220].


\bibitem{BenvenutiKruczenski05Ypq}
S. Benvenuti and M. Kruczenski,
``Semiclassical strings in Sasaki-Einstein manifolds and long operators in $\mathcal{N}=1$ gauge theories'',  
JHEP {\bf 10} (2006) 051, [hep-th/0505046].


\bibitem{KlebanovItzhaki02}
N. Itzhaki, I.R. Klebanov and S. Mukhi, 
``PP wave limit and enhanced supersymmetry in gauge theories'',  
JHEP {\bf 07} (2002) 044, [hep-th/0202153].


\bibitem{GomisOoguri02}
J. Gomis and H. Ooguri, 
``Penrose limit of $\mathcal{N}=1$ gauge theories'',  
Nucl. Phys. B {\bf 635} (2002) 106, [hep-th/0202157].


\bibitem{PandoZayasSonnenschein02}
L. A. Pando Zayas and J. Sonnenschein, 
``On Penrose limit and gauge theories'',  
JHEP {\bf 05} (2002) 010, [hep-th/0202186].


\bibitem{KupersteinSonnenschein06}
S. Kuperstein, O. Mintkevitch and J. Sonnenschein, 
``On the pp-wave limit and the BMN structure of the new Sasaki-Einstein spaces'',  
JHEP {\bf 12} (2006) 059, [hep-th/0609194].


\bibitem{MetsaevTseytlin98}
R. Metsaev and A. Tseytlin,
``Type IIB superstring action in $AdS_5 \times S^5$ background'',  
Nucl. Phys. B {\bf 533} (1998) 109, [hep-th/9805028].


\bibitem{BreitenlohnerFreedman}
P. Breitenlohner and D. Freedman,
``Positive energy in anti-de Sitter backgrounds and gauged extended supergravity'', 
Phys. Lett. B {\bf 115} (1982) 197.


\bibitem{GubserMitra02}
S. Gubser and I. Mitra,  
``Some interesting violations of the Breitenlohner-Freedman bound'', 
JHEP {\bf 03} (2002) 048, [hep-th/0108239].


\bibitem{BH2005}
  S. Benvenuti and A. Hanany,
  ``Conformal manifolds for the conifold and other toric field theories,''
  JHEP {\bf 0508}, 024 (2005)
  [hep-th/0502043].


\bibitem{FreundRubin}
P. Freund and M. Rubin, 
``Dynamics of dimensional reduction'',  
Phys. Lett. B {\bf 97} (1980) 233.


\bibitem{CandelasOssa89}
P. Candelas and X. de la Ossa,  
``Comments on conifolds'', 
Nucl. Phys. B {\bf 342} (1990) 246.


\bibitem{ppwave1}
R. Guven, 
``Plane wave limits and $T$ duality'',  
Phys. Lett. B {\bf 482} (2000) 255, [hep-th/0005061].


\bibitem{ppwave2}
M. Blau, J. Figueroa-O'Farrill, C. Hull and G. Papadopoulos, 
``Penrose limits and maximal supersymmetry'',  
Class. Quant. Grav. {\bf 19} (2002) L87, [hep-th/0201081];
``A new maximally supersymmetric background of IIB superstring theory'', 
JHEP {\bf 01} (2002) 047, [hep-th/0110242].


\bibitem{ppwave3}
M. Blau, J. Figueroa-O'Farrill and G. Papadapoulos, 
``Penrose limits, supergravity and brane dynamics'',  
Class. Quant. Grav. {\bf 19} (2002) 4753, [hep-th/0202111].



\bibitem{CahenWallach}
M. Callen and N. Wallach,
``Lorentzian symmetric spaces'', 
Bull. Am. Math. Soc. {\bf 76} (1970) 585.




\end{thebibliography}
\end{document}